\documentclass[conference,10pt]{IEEEtran}

\usepackage{tcolorbox}
\usepackage{algorithm}
\usepackage[noend]{algpseudocode}
\usepackage{bm}
\usepackage{stackrel}
\usepackage{subfigure}
\usepackage{epsf}
\usepackage{amsmath,amssymb}
\usepackage{graphicx}
\usepackage{color}
\usepackage{cite}
\usepackage{multirow,tabularx}
\usepackage{ifthen}
\usepackage{epstopdf}
\input{epstopdf.sty}
\usepackage{framed}
\usepackage{times}
\usepackage{pdfpages}

\usepackage{xr-hyper}
\input{epsf.sty}

\makeatletter
\def\BState{\State\hskip-\ALG@thistlm}
\makeatother

\newcommand{\newpart}[1]{{#1}}

\newcommand{\hide}[1]{\ifthenelse{\boolean{false}}{#1}{}}

\newtheorem{theorem}{{\bf Theorem}}

\newtheorem{lemma}{{\bf Lemma}}

\newtheorem{corollary}{{\bf Corollary}}

\newtheorem{definition}{{\bf Definition}}
\newtheorem{result}{{\bf Result}}

\newcommand{\qed}{\nobreak \ifvmode \relax \else
      \ifdim\lastskip<1.5em \hskip-\lastskip
      \hskip1.5em plus0em minus0.5em \fi \nobreak
      \vrule height0.75em width0.5em depth0.25em\fi}

\newcommand{\beq}{\begin{equation}}
\newcommand{\eeq}{\end{equation}}
\newcommand{\barr}{\begin{array}}
\newcommand{\earr}{\end{array}}

\newcommand{\benum}{\begin{enumerate}}
\newcommand{\eenum}{\end{enumerate}}

\newcommand{\bit}{\begin{itemize}}
\newcommand{\eit}{\end{itemize}}

\newcommand{\bc}{\begin{center}}
\newcommand{\ec}{\end{center}}

\newcommand{\bdes}{\begin{description}}
\newcommand{\edes}{\end{description}}

\newcommand{\bfig}{\begin{figure}}
\newcommand{\efig}{\end{figure}}

\newcommand{\bemq}{\begin{quote} \begin{em}}
\newcommand{\eemq}{\end{em} \end{quote}}

\newcommand{\bmp}{\begin{minipage}}
\newcommand{\emp}{\end{minipage}}

\newcommand{\brac}[1]{\left({#1}\right)}

\newcommand{\EX}[1]{\mathbb{E}\left[{#1}\right]} 

\newcommand{\prob}[1]{\text{Pr}\brac{#1}}

\newcommand{\bsp}{\begin{slide*}}
\newcommand{\esp}{\end{slide*}}
\newcommand{\bsl}{\begin{slide}}
\newcommand{\esl}{\end{slide}}

\newcommand{\blem}{\begin{lemma}}
\newcommand{\elem}{\end{lemma}}
\newcommand{\bthm}{\begin{theorem}}
\newcommand{\ethm}{\end{theorem}}

\newcommand{\pr}[1]{\mathbf{P}\left[ #1 \right]}

\IEEEoverridecommandlockouts

\begin{document}

\title{Optimizing Information Freshness in Wireless Networks under General Interference Constraints}
\date{\today}
\author{Rajat Talak, Sertac Karaman, Eytan Modiano
\thanks{This work was supported by NSF Grants AST-1547331, CNS-1713725, and
CNS-1701964, and by Army Research Office (ARO) grant number W911NF-
17-1-0508. This work was presented in part at ACM MobiHoc 2018, and received the best paper award at the conference~\cite{talak18_Mobihoc}. This paper has been accepted for publication in IEEE/ACM Transactions on Networking.}
\thanks{The authors are with the Laboratory for Information and Decision Systems (LIDS) at the Massachusetts Institute of Technology (MIT), Cambridge, MA. {\tt \{talak, sertac, modiano\}@mit.edu} }
}


\maketitle

\begin{abstract}
Age of information (AoI) is a recently proposed metric for measuring information freshness. AoI measures the time that elapsed since the last received update was generated. We consider the problem of minimizing average and peak AoI in a wireless networks, consisting of a set of source-destination links, under general interference constraints.
When fresh information is always available for transmission, we show that a stationary scheduling policy is peak age optimal. We also prove that this policy achieves average age that is within a factor of two of the optimal average age. In the case where fresh information is not always available, and packet/information generation rate has to be controlled along with scheduling links for transmission, we prove an important \emph{separation principle}: the optimal scheduling policy can be designed assuming fresh information, and independently, the packet generation rate control can be done by ignoring interference. Peak and average AoI for discrete time G/Ber/1 queue is analyzed for the first time, which may be of independent interest.
\end{abstract}


\section{Introduction}
\label{sec:intro}
Exchanging status updates, in a timely fashion, is an important functionality in many network settings. In unmanned aerial vehicular (UAV) networks, exchanging position, velocity, and control information in real time is critical to safety and collision avoidance~\cite{FANETs2013, talakCDC16}. In internet of things (IoT) and cyber-physical systems, information updates need to be sent to a common ground station in a timely fashion for better system performance~\cite{KimKumar2012_CPSperspective}. In cellular networks, timely feedback of the link state information to the mobile nodes, by the base station, is necessary to perform opportunistic scheduling and rate adaptation~\cite{GuharoyMehta2013_FeedbackDelay, LTE_book}.

Traditional performance measures, such as delay or throughput, are inadequate to measure the timeliness of the updates, because delay or throughput are packet centric measures that fail to capture the timeliness of the information from an application perspective. For example, a packet containing stale information is of little value even if it is delivered promptly by the network. In contrast, a packet containing freshly updated information may be of much greater value to the application, even if it is slightly delayed.

\begin{figure}
  \centering
  \includegraphics[width=0.47\textwidth]{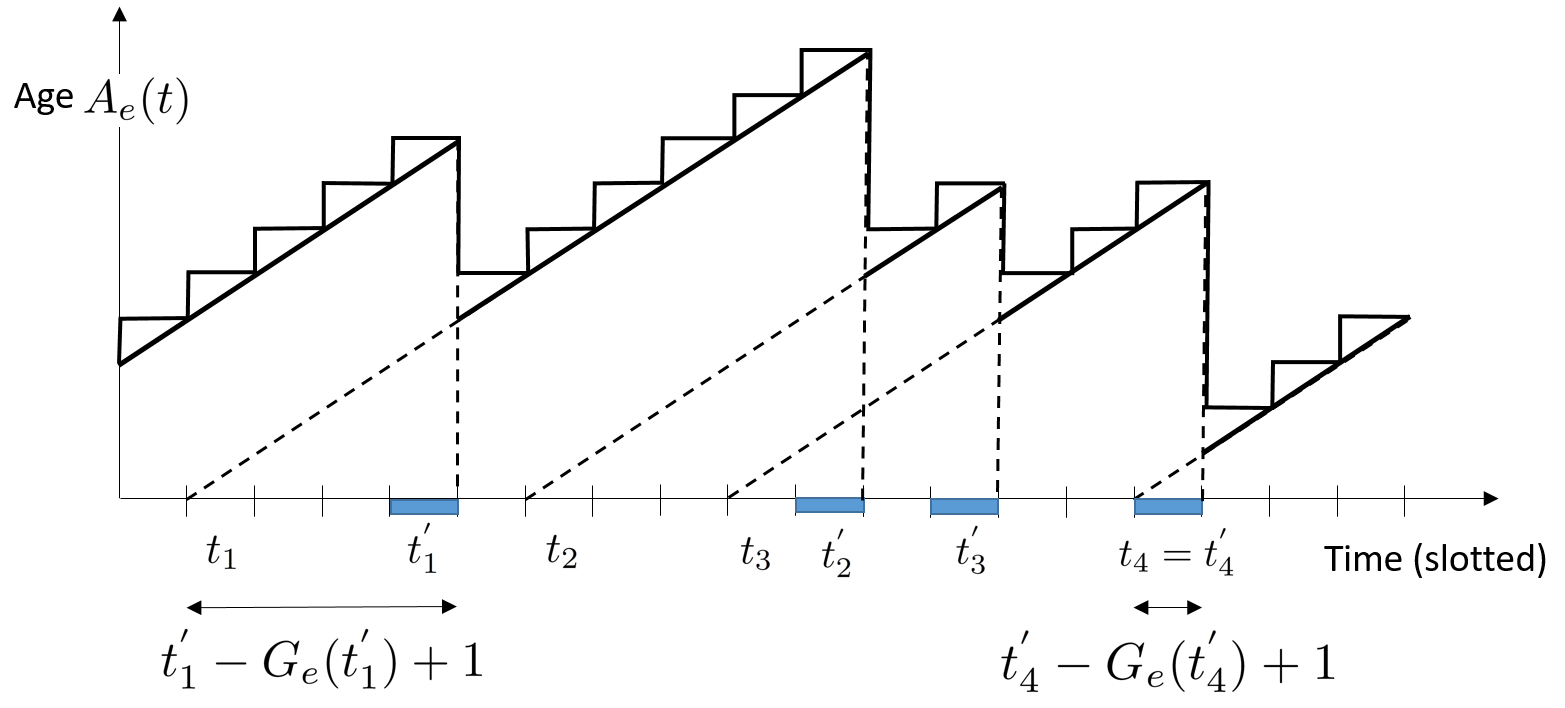} 
  \caption{Time evolution of age, $A_{e}(t)$, of a link $e$. Times $t_i$ and $t^{'}_i$ are instances of $i$th packet generation and reception, respectively. Given the definition $G_{e}(t^{'}_{i}) \triangleq t_{i}$, the age is reset to $t^{'}_i - G_{e}(t^{'}_{i}) + 1$ when the $i$th packet is received.}
  \label{fig:age}
\end{figure}

A new measure, called Age of Information (AoI), was proposed in~\cite{2011SeCON_Kaul, 2012Infocom_KaulYates} that measures the time that elapsed since the last received update was generated. Figure~\ref{fig:age} shows evolution of AoI for a destination node as a function of time. The AoI, upon reception of a new update packet, drops to the time elapsed since generation of this packet, and grows linearly otherwise. AoI being a destination-node centric measure, rather than a packet centric measure like throughput or delay, is more appropriate to measure timeliness of updates.

In~\cite{2011SeCON_Kaul}, AoI was first studied for a vehicular network using simulations. Nodes generated fresh update packets periodically at a certain rate, which were queued at the MAC layer first-in-first-out (FIFO) queue for transmission. An optimal packet generation rate was observed that minimized age. It was further observed that the age could be improved by controlling the MAC layer queue, namely, by limiting the buffer size or by changing the queueing discipline to last-in-first-out (LIFO). However, the MAC layer queue may not be controllable in practice. This lead to several works on AoI under differing assumptions on the ability to control the MAC layer queue.

Many of the applications where age is an important metric involve wireless networks, and interference constraints are one of the primary limitations to system performance. However, theoretical understanding of age of information under interference constraints has received little attention thus far. In~\cite{2016Ep_WiOpt}, the problem of scheduling finite number of update packets under physical interference constraint for age minimization was shown to be NP-hard. Age for a broadcast network, where only a single link can be activated at any time, was studied in~\cite{2016allerton_IgorAge, 2017ISIT_YuPin}, and preliminary analysis of age for a slotted ALOHA-like random access was done in~\cite{2017X_KaulYates_AoI_ALOHA}.

We consider the problem of minimizing age of information in wireless networks under general interference constraints, and time-varying links. The wireless network consists of a set of source-destination pairs, \newpart{each} connected by a wireless link. Each source generates information updates, which are to be sent to its destination. We measure the age at each destination. We consider average age, which is the time average of the age curve in Figure~\ref{fig:age}, and peak age, which is the average of all the peaks in the age curve in Figure~\ref{fig:age}, as the \newpart{metrics} of performance. Due to wireless interference only a subset of links can be made to transmit simultaneously. We obtain simple scheduling policies that are optimal, or nearly optimal.

We consider two types of sources: \emph{active sources} and \emph{buffered sources}. Active sources can generate a new update packet for every transmission, i.e., fresh information is always available for transmission. Buffered sources, on the other hand, can only control the rate of packet generation, while the generated packets are buffered in the MAC layer FIFO queue for transmission.

For a network with active sources, we show that a stationary scheduling policy, in which links are activated according to a stationary probability distribution, is peak age optimal. We also show that this policy achieves average age that is within factor of two of the optimal average age. Moreover, we prove that this optimal policy can be obtained as a solution to a convex optimization problem.

For a network with buffered sources, we consider Bernoulli and periodic packet generation, that generate update packets at a certain rate. We design a rate control and scheduling policy to minimizes age. We show that if rate control is performed assuming that there is no other link in the network, and scheduling is done in the same way as in the active source case, then this is close to the optimal age achieved by jointly minimizing over stationary scheduling policies and rate control.
This \emph{separation principle} provides an useful insight towards the design of age optimal policies, as scheduling and rate control are typically done at different layers of the protocol stack.
%

Peak and average age for the discrete time FIFO G/Ber/1 queue is analyzed for the first time, which may be of independent interest. A preliminary version of this work appeared in MobiHoc 2018~\cite{talak18_Mobihoc}.

\subsection{Literature Survey}
Motivated by~\cite{2011SeCON_Kaul}, AoI was first studied for the first come first \newpart{served} (FCFS) M/M/1, M/D/1, and D/M/1 queues in~\cite{2012Infocom_KaulYates}. Since then, AoI has been analyzed for several queueing systems~\cite{2012Infocom_KaulYates, 2012CISS_KaulYates, 2013ISIT_KamKomEp, 2014ISIT_KamKomEp, 2014ISIT_CostaEp, 2015ISIT_LongBoEM, 2016X_Najm, sun_lcfs_better, Inoue17_FCFS_AoIDist, 2018_Ulukus_GG11, 2018ISIT_Yates_AoI_ParallelLCFS, 2011SeCON_Kaul, 2016_MILCOM_Ep_AoI_Buffer_Deadline_Replace, 2016_ISIT_Ep_AoI_Deadlines, 2018_ISIT_Inoue_AoI_Deadline}, with the goal to minimize AoI. Two time average metrics of AoI, namely, peak and average age are generally considered. AoI for multiclass M/G/1 and G/G/1 queues, under FIFO service, were  studied in~\cite{2015ISIT_LongBoEM}. Age for a M/M/$\infty$ was analyzed in~\cite{2013ISIT_KamKomEp}, which studied the impact of out-of-order delivery of packets on age, while the effect of packet errors or packet drops on age was studied in~\cite{2016X_LongBo}. Age for LIFO queues was analyzed under various arrival and service time distributions in~\cite{2012CISS_KaulYates, 2016X_Najm, 2014ISIT_CostaEp}.

The advantage of having parallel servers, towards improving AoI, was demonstrated in~\cite{2014ISIT_KamKomEp, 2014ISIT_CostaEp, 2018ISIT_Yates_AoI_ParallelLCFS}. Having smaller buffer sizes~\cite{2011SeCON_Kaul, 2016_MILCOM_Ep_AoI_Buffer_Deadline_Replace} or introducing packet deadlines~\cite{2016_MILCOM_Ep_AoI_Buffer_Deadline_Replace, 2016_ISIT_Ep_AoI_Deadlines, 2018_ISIT_Inoue_AoI_Deadline}, in which a packet deletes itself after its deadline expiration, are two other considered ways of improving AoI. In~\cite{sun_lcfs_better}, the LCFS queue scheduling discipline, with preemptive service, is shown to be an age optimal, when the service times are exponentially distributed, \newpart{while in~\cite{yin17_tit_update_or_wait} optimal update generation policy to improve age is investigated.}

AoI for energy harvesting communication systems was considered in~\cite{2015ISIT_Yates_AoIEH_LazyTimely, 2015ITA_Elif_AoIEH_Replenishment_Constraints, 2017_GC_Ulukus_AoIEH_TwoHop, 2017ISIT_Elif_AoIEHScheduling, 2018TGreenCN_AoIEH, 2018TIT_UlukusPoor_AoIEH_OnlinePolicies}, while AoI for gossip type information dissemination was analyzed in~\cite{2009AoGossip_MeanField, 2013AoGossip}.

Very little work existed, prior to this, on link scheduling to minimize AoI. Scheduling, for AoI minimization, finitely many packets under physical interference constraints was shown to be NP-hard in~\cite{2016Ep_WiOpt}. Index policies were proposed in~\cite{2016allerton_IgorAge, 2017ISIT_YuPin} for broadcast network, and AoI minimization for slotted ALOHA-like random access was considered in~\cite{2017X_KaulYates_AoI_ALOHA}. In this work, we consider the problem of age minimization for wireless networks, under general interference constraints. We consider single-hop information flows.

This paper also introduces a distinction between active sources and buffered sources, and shows that \newpart{an} optimal scheduling policy in one case can be nearly optimal in the other. A preliminary version of this work appeared in~\cite{talak18_Mobihoc}, and several extensions have recently appeared in~\cite{talak17_allerton, talak18_greece, talak18_ISIT, talak18_WiOpt}. In~\cite{talak18_greece}, we derive distributed policies for age minimization, while in~\cite{talak18_WiOpt} and~\cite{talak18_ISIT} we propose age-based and a virtual queue based policy for age minimization. We show in~\cite{talak18_WiOpt}, that using the current channel state information can result in significant age improvement. The case of multi-hop information flows was considered in~\cite{talak17_allerton}.

\subsection{Organization}

The rest of this paper is organized as follows. We describe the system model in Section~\ref{sec:model}. Age minimization for active sources is considered in Section~\ref{sec:active_source}, where we also characterize the stationary policy that minimizes peak age under a general interference model. Age minimization for buffered sources is discussed in Section~\ref{sec:queued} and Section~\ref{sec:non-stationary}. Numerical results are presented in Section~\ref{sec:num}, and we conclude in Section~\ref{sec:conclusion}.

\section{System Model}
\label{sec:model}
We consider a wireless communication network as a graph $G = (V,E)$, where $V$ is the set of nodes and $E$ is the set of communication links between the nodes in the network. Each communication link $e \in E$ is a source-destination pair in the network. The source generates information updates that need to be communicated to the destination.
Time is slotted and the duration of each slot is normalized to unity.

Wireless interference constraints limit the set of links which can be activated simultaneously~\cite{ak_winet}. 
We call a set $m \subset E$ to be a \emph{feasible activation set} if all links in $m$ can be activated simultaneously without interference, and denote by $\mathcal{A}$ the collection of all feasible activation sets. We call this the \emph{general interference model}, as it incorporates several popular interference models such as 1-hop interference, \newpart{$k$-hop} interference~\cite{gaurav_ravi_ness2006_scheduling_complexity}, and protocol interference models~\cite{ak_winet}.

A non-interfering transmission over link $e$ does not always succeed due to channel errors. We let $R_{e}(t) \in \{1,0\}$ denote the channel error process for link $e$, where $R_{e}(t) = 1$ if a non-interfering transmission over link $e$ succeeds and $R_{e}(t) = 0$ otherwise. We assume $R_{e}(t)$ to be independent across links, and i.i.d. across time with $\gamma_e = \pr{R_{e}(t) = 1} > 0$, for all $e \in E$. We assume that the channel success probabilities $\gamma_e$ are known, or can be measured separately.

We consider two types of sources, namely, \emph{active source} and \emph{buffered source}. An active source can generate a new update packet at the beginning of each slot for transmission, while discarding old update packets that were not transmitted. Thus, for an active source, a transmitted packet always contains fresh information. Packets generated by a buffered source, on the other hand, get queued before transmission, and may contain `stale' information. The source cannot control this FIFO queue, and thus, the update packets have to incur queueing delay. A buffered source, however, can control the packet generation rate.

The age $A_{e}(t)$ at the destination of link $e$ evolves as shown in Figure~\ref{fig:age}. When the link $e$ is activated successfully in a slot, the age $A_{e}(t)$ is reduced to the time elapsed since the generation of the delivered packet. $A_{e}(t)$ grows linearly in absence of any communication over $e$. This evolution can be simply described as
\begin{equation}
\label{eq:age_update}
A_{e}(t+1) = \left\{\begin{array}{ll}
                       \!\! t - G_{e}(t) + 1     &\!\! \text{if $e$ is activated at $t$} \\
                       \!\! A_{e}(t) + 1 &\!\! \text{if $e$ is not activated at $t$}
                     \end{array}\right.,
\end{equation}
where $G_{e}(t)$ is the generation time of the packet delivered over link $e$ at time $t$. In the active source case, for example, $G_{e}(t) = t$ since a new update packet is made available at the beginning of each slot. Thus, in this case, the age $A_{e}(t)$ is equal to the time elapsed since an update packet was transmitted over it, i.e., the last \emph{activation} of link $e$. For ease of presentation, we will refer to $A_{e}(t)$ as the age of link $e$.

We define two metrics to measure long term age performance over a network of interfering links. The \emph{weighted average age}, given by,
\begin{equation}
\label{eq:ave_age}
A^{\text{ave}} = \limsup_{T \rightarrow \infty} \frac{1}{T} \sum_{t=1}^{T} \sum_{e \in E} w_e A_{e}(t),
\end{equation}
where $w_e$ are positive weights denoting the relative importance of each link $e \in E$, and the \emph{weighted peak age}, given by,
\begin{equation}
\label{eq:peak_age}
A^{\text{p}} = \limsup_{N \rightarrow \infty} \frac{1}{N}\sum_{i = 1}^{N}\sum_{e \in E} w_e  A_{e}\left( T_{e}(i)\right),
\end{equation}
where $T_{e}(i)$ denotes the time at which link $e$ was successfully activated for the $i$th time. Peak age is the average of age peaks, which happen just before link activations. Without loss of generality we assume that $w_e > 0$ for all $e$, and the link weights are normalized to sum to unity, i.e., $\sum_{e \in E} w_e = 1$.



\subsection{Scheduling Policies}

A scheduling policy is needed in order to decide which links to activate at any time slot. It determines the set of links $m(t) \subset E$ that will be activated at each time $t$. The policy can make use of the past history of link activations and age to make this decision, i.e., at each time $t$ the policy $\pi$ will determine $m(t)$ as a function of the set
\begin{equation}
\label{eq:history}
\mathcal{H}(t) = \{ m(\tau), \mathbf{R}(\tau), \mathbf{A}(\tau') | 0 \leq \tau \leq t-1~\text{and}~0 \leq \tau' \leq t \},
\end{equation}
where $\mathbf{A}(t) = \left( A_{e}(t)\right)_{e \in E}$ and $\mathbf{R}(t) = \left( R_{e}(t)\right)_{e \in E}$. Note that $\mathbf{R}(t) \notin \mathcal{H}(t)$, i.e., the current channel state $\mathbf{R}(t)$ is not observed before making \newpart{a} decision at time $t$. 
We consider centralized scheduling policies, in which this information is centrally available to a scheduler, which is able to implement its scheduling decisions.

Given such a policy $\pi$, define the \emph{link activation frequency} $f_{e}(\pi)$, for a link $e$, to be the fraction of times link $e$ is successfully activated, i.e.,
\begin{equation}
f_{e}(\pi) = \lim_{T \rightarrow \infty} \frac{\sum_{t=1}^{T}\mathbb{I}_{\left\{e \in m(t), m(t) \in \mathcal{A} \right\}}}{T}, \label{eq:link_act_freq}
\end{equation}
where $m(t)$ is the set of links activated at time $t$, and $\mathbb{I}_{S}$ is an indicator function which equals $1$ only if $S$ holds, and $0$ otherwise. Note that $f_{e}(\pi)$ is not the frequency of successful activations, as channel errors can render an activation of a link unsuccessful.
%
If $f_{e}(\pi) = 0$ for a certain link $e$ then the average and peak age will be unbounded. We, therefore, limit our attention to the set of policies $\Pi$ for which $f_{e}(\pi)$ is well defined and strictly positive for all $e \in E$:
\begin{equation}\label{eq:Pi}
\Pi = \left\{ \pi \big| f_{e}(\pi)~\text{exists and}~f_{e}(\pi) > 0~\forall e \in E\right\}.
\end{equation}

We define the set of all feasible link activation frequencies, for policies described above:
\begin{equation}
\mathcal{F} = \left\{ \mathbf{f} \in \mathbb{R}^{|E|}~|~f_e = f_e(\pi)~\forall~e \in E~\text{and some}~\pi \in \Pi \right\}. \nonumber
\end{equation}
This set can be characterized by linear constraints as
\begin{equation}
\label{eq:mathcalF}
\mathcal{F} = \left\{ \mathbf{f} \in \mathbb{R}^{|E|}~|~\mathbf{f} = M \mathbf{x},~\mathbf{1}^{T}\mathbf{x} \leq 1~\text{and}~\mathbf{x} \geq 0 \right\},
\end{equation}
where $\mathbf{x}$ is a vector in $\mathbb{R}^{|\mathcal{A}|}$ and $M$ is a $|E|\times |\mathcal{A}|$ matrix with elements
\begin{equation}
\label{eq:M}
M_{e,m} = \left\{ \begin{array}{ll}
                    1 &~\text{if}~e \in m \\
                    0 &~\text{otherwise}
                  \end{array}\right.,
\end{equation}
for all links $e$ and feasible activation sets $m \in \mathcal{A}$.

A simple sub-class of policies, which do not use any past history, is the class of stationary policies. In it, links are activated independently across time according to a stationary distribution. We define a stationary policy as follows:
\begin{definition}[(Randomized) Stationary Policies]
Let $B_{e}(t) = \{ e \in m(t), m(t) \in \mathcal{A}\}$ be the event that link $e$ was activated at time $t$. Then, the policy $\pi$ is stationary if
\begin{enumerate}
    \item $B_{e}(t)$ is independent across $t$, and
    \item $\prob{B_{e}(t_1)} = \prob{B_{e}(t_2)}$ for all $t_1, t_2 \in \left\{1, 2, \ldots\right\}$,
\end{enumerate}
for all $e \in E$. For the ease of presentation, we shall refer to randomized stationary policies and stationary policies.  We use $\Pi_{\text{st}}$ to denote the space of all stationary policies in the policy space $\Pi$.
\end{definition}

The stationary policies defined above are also memoryless, in the sense that $B_{e}(t)$ are independent across time. The following are two examples of stationary policies.

\emph{Example~1}: Set $p_e \in (0,1)$ for all $e \in E$, and let a policy attempt transmission over link $e$ with probability $p_e$, independent of other link's attempts. These are distributed policies, and are investigated in~\cite{talak18_greece}.

\emph{Example~2}: Assign a probability distribution $\mathbf{x} \in \mathbb{R}^{|\mathcal{A}|}$ over the collection of feasible activation sets, $\mathcal{A}$. Then, in each slot, activate the set $m \in \mathcal{A}$ with probability $x_m$, independent across time. We call this the \emph{stationary centralized policy}.
For this policy,
\begin{equation}
\prob{B_{e}(t)} = \sum_{m: e \in m} x_m = \left( M\mathbf{x} \right)_{e},
\end{equation}
for all $e \in E$ and slots $t$.

We will see in the next section that in the active source case, a stationary centralized policy is peak age optimal, and is within a factor of $2$ from the optimal average age, over the space $\Pi$. Motivated by this, in Section~\ref{sec:queued}, we will consider stationary scheduling policies for the buffered sources.

\subsection{Rate Control in Buffered Sources}
In the case of active sources, the update packets are generated at every transmission opportunity. However, in the case of buffered sources, we can only control the update generation rate. We consider two models of update generation, namely, Bernoulli update generation and periodic update generation. In the Bernoulli update generation, the source of link $e$ generates a new update in a time slot with probability $\lambda_e$. In the periodic update generation, the source of link $e$ generates a new update once every $D_e = 1/\lambda_e$ slots.

We consider the problem of jointly determining the update generation rate $\lambda_e$, for each link, and the stationary scheduling policy $\pi \in \Pi_{\text{st}}$, in order to minimize peak and average age defined in~\eqref{eq:peak_age} and~\eqref{eq:ave_age}, respectively. In Section~\ref{sec:queued}, We propose a \emph{separation principle policy} that performs very close to the optimal.


\section{Minimizing Age with Active Sources}
\label{sec:active_source}
We consider a network where all the sources are active. Since the age metrics depend on the policy $\pi \in \Pi$ used, we make this dependence explicit by the notation $A^{\text{ave}}(\pi)$ and $A^{\text{p}}(\pi)$. We use $A^{\text{ave}\ast}$ and $A^{\text{p}\ast}$ to denote the minimum average and peak age, respectively, over all policies in $\Pi$.

We first characterize the peak age for any policy $\pi \in \Pi$, and show that a stationary centralized policy is peak age optimal.
\begin{framed}
\begin{theorem}
\label{thm:peak_age_char}
For any policy $\pi \in \Pi$, the peak age is given by
\begin{equation}
\label{eq:local1}
A^{\text{p}}(\pi) = \sum_{e \in E} \frac{w_e}{\gamma_e f_e\left( \pi \right)}.
\end{equation}
As a consequence, for every $\pi \in \Pi$ there exists a stationary policy $\pi_{\text{st}} \in \Pi_{\text{st}}$ such that $A^{\text{p}}(\pi) = A^{\text{p}}(\pi_{\text{st}})$. Thus, a stationary policy is peak age optimal.
\end{theorem}
\end{framed}
\begin{IEEEproof}
Consider a randomized stationary policy with link activation frequency $f_e = \pr{B_{e}(t)}$. \newpart{It activates link $e$ with probability $f_e$, in each time slot. Since the channel process is also i.i.d., it follows that the link $e$ is successfully activated in a slot with probability $f_e \gamma_e$.}

\newpart{This implies that the inter-(successful) activation time of link $e$ is geometrically distributed with mean $\frac{1}{\gamma_e f_e}$.} Therefore, the peak age $A^{\text{p}}_e$, which is nothing but the average inter-(successful) activation time, equals $\frac{1}{\gamma_e f_e}$, i.e. $A^{\text{p}}_e = \frac{1}{\gamma_e f_e}$.

Thus, the weighted peak age is $\sum_{e \in E}\frac{w_e}{\gamma_e f_e}$. The same result extends to any policy $\pi \in \Pi$, due to the existence of the limit~\eqref{eq:link_act_freq}, which ensures ergodicity of the link activation process $U_{e}(t) = \mathbb{I}_{\{e \in m(t), m(t) \in \mathcal{A}\}}$. The detailed arguments are presented in Appendix~\ref{pf:thm:peak_age_char}.
\end{IEEEproof}
Theorem~\ref{thm:peak_age_char} implies that the peak age minimization problem can be written as
\begin{align}\label{eq:opt_peak_age_problem0}
\begin{aligned}
& \underset{\mathbf{f}}{\text{Minimize}}
& & \sum_{e \in E} \frac{w_e}{\gamma_e f_e} \\
& \text{subject to} & & \mathbf{f} \in \mathcal{F},
\end{aligned}
\end{align}
where $\mathcal{F}$ - given in~\eqref{eq:mathcalF} - is the space of all link activation frequencies for policies in $\Pi$. We discuss solutions to~\eqref{eq:opt_peak_age_problem0} under general, and more specific, interference constraints in Section~\ref{sec:centralized}.

Theorem~\ref{thm:peak_age_char} implies that a stationary centralized policy is peak age optimal. We will next show that a peak age optimal stationary policy is also within a factor of $2$ from the optimal average age. We first show an important relation between peak and average age for any policy $\pi \in \Pi$.
\begin{framed}
\begin{theorem}
\label{thm:age_relation}
For all $\pi \in \Pi$ we have
\begin{equation} \label{eq:tempsss}
A^{\text{p}}(\pi) \leq 2A^{\text{ave}}(\pi) - 1.
\end{equation}
\end{theorem}
\end{framed}
\begin{IEEEproof}
The result is a direct implication of Cauchy-Schwartz inequality. See Appendix~\ref{pf:thm:age_relation}.
\end{IEEEproof}
%

Let $A^{\text{p}\ast} = \min_{\pi \in \Pi} A^{\text{p}}(\pi)$ and $A^{\text{ave}\ast} = \min_{\pi \in \Pi} A^{\text{ave}}(\pi)$ be the optimal peak and average age, respectively, over the space of all policies in $\Pi$. Since the relation~\eqref{eq:tempsss} holds for every policy $\pi \in \Pi$, it is natural to expect it to hold at the optimality. This is indeed true.
\begin{framed}
\begin{corollary}
\label{cor:age_relation}
The optimal peak age is bounded by
\begin{equation}
A^{\text{p}\ast} \leq 2 A^{\text{ave}\ast} - 1.
\end{equation}
\end{corollary}
\end{framed}
\begin{IEEEproof}
Since $A^{\text{p}\ast}$ is the optimal peak age we have $A^{\text{p}\ast} \leq A^{\text{p}}(\pi)$ for any policy $\pi$. Substituting this in~\eqref{eq:tempsss} we get $A^{\text{p}\ast} \leq 2 A^{\text{ave}}(\pi) - 1$, for all $\pi \in \Pi$. Minimizing the right hand side over all $\pi \in \Pi$ we obtain the result.
\end{IEEEproof}

We now show that for any stationary policy the average and peak age are equal.
\begin{framed}
\begin{lemma}
\label{lem:rp}
We have $A^{\text{ave}}(\pi) = A^{\text{p}}(\pi)$ for any stationary policy $\pi \in \Pi_{\text{st}}$.
\end{lemma}
\end{framed}
\begin{IEEEproof}
Let $S_e$ be the time between two successful activations of link $e$, under the stationary policy $\pi \in \Pi_{\text{st}}$. We show that the peak age and average age, for a link $e$, is given by $A^{\text{p}}_{e}(\pi) = \EX{S_e}$ and $A^{\text{ave}}_e(\pi) = \frac{\EX{S^{2}_{e}}}{2\EX{S_e}} + \frac{1}{2}$, respectively. Then, noting the fact that $S_e$ is a geometrically distributed random variable, for $\pi \in \Pi_{\text{st}}$, we get the result. The detailed proof is given in Appendix~\ref{pf:lem:rp}.
\end{IEEEproof}

An immediate implication of Corollary~\ref{cor:age_relation} and Lemma~\ref{lem:rp} is that a stationary peak age optimal policy is also within a factor of $2$ from the optimal average age.
\begin{framed}
\begin{theorem}
\label{thm:peak_ave}
If $\pi_C$ is a stationary policy that minimizes peak age over the policy space $\Pi$ then the average age for $\pi_C$ is within factor $2$ of the optimal average age. Specifically,
\begin{equation}
A^{\text{ave}\ast} \leq A^{\text{ave}}(\pi_C) \leq 2 A^{\text{ave}\ast} - 1.
\end{equation}
\end{theorem}
\end{framed}
\begin{IEEEproof}
Let $\pi_C$ be the stationary policy that minimizes peak age. We, thus, have
\begin{equation}\label{eq:ww1}
A^{\text{p}}(\pi_C) = A^{\text{p}\ast},
\end{equation}
Since $\pi_C$ is also a stationary policy, Lemma~\ref{lem:rp} implies
\begin{equation} \label{eq:ww2}
A^{\text{ave}}(\pi_C) = A^{\text{p}}(\pi_C).
\end{equation}
Using \eqref{eq:ww1},~\eqref{eq:ww2}, and Corollary~\ref{cor:age_relation} we obtain
\begin{equation}\label{eq:n1}
A^{\text{ave}}(\pi_C) = A^{\text{p}}(\pi_C) = A^{\text{p}\ast} \leq 2 A^{\text{ave}\ast} - 1.
\end{equation}
This proves the result.
\end{IEEEproof}

Theorem~\ref{thm:peak_ave} tells us that the stationary peak age optimal policy obtained by solving~\eqref{eq:opt_peak_age_problem0} is within a factor of $2$ of optimal average age. Motivated by this, we next characterize solutions to the problem~\eqref{eq:opt_peak_age_problem0}.

\subsection{Optimal Stationary Policy $\pi_C$}
\label{sec:centralized}
The peak age minimization problem~\eqref{eq:opt_peak_age_problem0} over $\mathcal{F}$ can be written as
\begin{align}\label{eq:opt_peak_age_problem}
\begin{aligned}
& \underset{\mathbf{x} \in \mathbb{R}^{|\mathcal{A}|}}{\text{Minimize}}
& & \sum_{e \in E} \frac{w_e}{\gamma_e f_e} \\
& \text{subject to}
& & \mathbf{f} = M\mathbf{x} \\
&&& \mathbf{1}^{T}\mathbf{x} \leq 1,~\mathbf{x} \geq 0
\end{aligned}
\end{align}
Note that the optimization is over $\mathbf{x}$, the activation probabilities of feasible activation sets $m \in \mathcal{A}$. This is because the link activation frequencies $\mathbf{f}$ get completely determined by $\mathbf{x}$.
The problem~\eqref{eq:opt_peak_age_problem} is a convex optimization problem in standard form~\cite{boyd}. The solution to it is a vector $\mathbf{x} \in \mathbb{R}^{|\mathcal{A}|}$ that defines a probability distribution over link activation sets $\mathcal{A}$, and determines a stationary centralized policy that minimizes peak age. Average age for this policy, by Theorem~\ref{thm:peak_ave}, is also within a factor of $2$ from the optimal average age. We denote this stationary centralized policy by $\pi_{C}$.

We first characterize the optimal solution to~\eqref{eq:opt_peak_age_problem} for any $\mathcal{A}$. Given $\mathbf{x} \in \mathbb{R}^{|\mathcal{A}|}$, a probability distribution over the link activation sets $\mathcal{A}$, $\mathbf{f} = M\mathbf{x} \in \mathbb{R}^{|E|}$ is the vector of induced link activation frequencies. Now, define $\Omega_{m}(\mathbf{x})$-weight for every  feasible link activation set $m \in \mathcal{A}$ as
\begin{equation}
\Omega_{m}(\mathbf{x}) = \sum_{e \in m} \frac{w_e}{\gamma_e \left(M\mathbf{x}\right)^{2}_{e}} = \sum_{e \in m} \frac{w_e}{\gamma_e f^{2}_e}.
\end{equation}
Clearly, $\Omega_{m}(\mathbf{x}) > 0$ for every $m$. We now characterize the optimal solution to~\eqref{eq:opt_peak_age_problem} in terms of $\Omega_{m}(\mathbf{x})$-weights.
\begin{framed}
\begin{theorem}
\label{thm:opt_peak_age_char}
$\mathbf{x} \in \mathbb{R}^{|\mathcal{A}|}$ solves~\eqref{eq:opt_peak_age_problem} if and only if there exists a $\Omega > 0$ such that
\begin{enumerate}
\item For all $m \in \mathcal{A}$ such that $x_{m} > 0$ we have $\Omega_{m}(\mathbf{x}) = \Omega$
\item $x_{m} = 0$ implies $\Omega_{m}(\mathbf{x}) \leq \Omega$
\item $\sum_{m \in \mathcal{A}} x_m = 1$ and $x_m \geq 0$
\end{enumerate}
Further, $\Omega$ is the optimal peak age $A^{\text{p}\ast}$.
\end{theorem}
\end{framed}
\begin{IEEEproof}
The problem~\eqref{eq:opt_peak_age_problem} is convex and Slater's conditions are trivially satisfied as all constraints are affine~\cite{boyd}. Thus, the KKT conditions are both necessary and sufficient. We use the KKT conditions to derive the result. See Appendix~\ref{pf:thm:opt_peak_age_char} for a detailed proof.
\end{IEEEproof}
Theorem~\ref{thm:opt_peak_age_char} implies that at the optimal distribution $\mathbf{x}$, all $m \in \mathcal{A}$ with positive probability, $x_m > 0$, have equal $\Omega_{m}(\mathbf{x})$-weights, while all other $m \in \mathcal{A}$ have smaller $\Omega_{m}(\mathbf{x})$-weights.

Although the set $\mathcal{A}$ is very large, it is mostly the case that only a small subset of it is assigned positive probability.
A feasible activation set $m$ is said to be \emph{maximal} if adding any link $e \notin m$, to $m$, renders it infeasible, i.e., $m \cup \{ e\} \notin \mathcal{A}$.
In the following we show that only the maximal sets in $\mathcal{A}$ are assigned positive probability, thereby reducing the number of constraints in~\eqref{eq:opt_peak_age_problem}.
\begin{framed}
\begin{corollary}\label{cor:maximal}
If $\mathbf{x}$ is the optimal solution to~\eqref{eq:opt_peak_age_problem} then $x_m = 0$ for all non-maximal sets $m \in \mathcal{A}$.
\end{corollary}
\end{framed}
\begin{IEEEproof}
Let $m' \in \mathcal{A}$ be a non-maximal set. Thus, there exists a $\overline{m} \in \mathcal{A}$ such that $m' \subsetneq \overline{m}$. By definition of $\Omega_{m}(\mathbf{x})$ we have $\Omega_{m'}(\mathbf{x}) < \Omega_{\overline{m}}(\mathbf{x})$. Thus, if $x_{m'} > 0$ then we would have $\Omega = \Omega_{m'}(\mathbf{x}) < \Omega_{\overline{m}}(\mathbf{x})$ which is a contradiction.
\end{IEEEproof}

The optimization problem~\eqref{eq:opt_peak_age_problem}, although convex, has a variable space that is $|\mathcal{A}|$-dimensional, and thus, its computational complexity increases exponentially in $|V|$ and $|E|$. It is, however, possible to obtain the solution efficiently in certain specific cases. 

\subsubsection{Single-Hop Interference Network}
\label{sec:matching}
Consider a network $G = (V, E)$ where links interfere with one another if they share a node, i.e., if they are adjacent. For this network, every feasible activation set is a matching on $G$, and therefore, $\mathcal{A}$ is a collection of all matchings in $G$. As a result, the constraint set in~\eqref{eq:opt_peak_age_problem} is equal to the matching polytope~\cite{comb_opt_book}. The problem of finding an optimal schedule reduces to solving a convex optimization problem~\eqref{eq:opt_peak_age_problem} over a matching polytope. This can be efficiently solved (i.e., in polynomial time) by using the Frank-Wolfe algorithm~\cite{Garber16_FrankWolfe}, and the separation oracle for matching polytope developed in~\cite{Hajek_Sasaki_OptScheduling}.

\subsubsection{$K$-Link Activation Network}
\label{sec:anyK}
Consider a network $G = (V, E)$ in which at most $K$ links can be activated at any given time; we label links $E = \{0, 2, \ldots |E|-1\}$. Such interference constraints arise in cellular systems where the $K$ represents the number of OFDM sub-channels or number of sub-frames available for transmission in a cell~\cite{LTE_book}.

The set $\mathcal{A}$ is a collection of all subsets of $E$ of size at most $K$. This forms a uniform matroid over $E$~\cite{comb_opt_book}. As a result, the constraint set in~\eqref{eq:opt_peak_age_problem} is the uniform matroid polytope.
It is known that the inequalities $\sum_{e \in E} f_e \leq K$ and $0\leq f_e \leq 1$, for all $e \in E$, are necessary and sufficient to describe this polytope~\cite{comb_opt_book}.
Thus, the peak age minimization problem~\eqref{eq:opt_peak_age_problem} reduces to
\begin{align}\label{eq:opt_klink}
\begin{aligned}
& \underset{\mathbf{f} \in [0,1]^{|E|}}{\text{Minimize}}
& & \sum_{e \in E} \frac{w_e}{\gamma_e f_e} \\
& \text{subject to}
& & \sum_{e \in E} f_e \leq K
\end{aligned}
\end{align}
Since the number of constraints is now linear in $|E|$, this problem can be solved using standard convex optimization algorithms~\cite{boyd}. 


\section{Minimizing Age with Buffered Sources}
\label{sec:queued}
We now consider a network with buffered sources, where each source generates update packets according to a Bernoulli process. The generated packets get queued at the MAC layer FIFO queue for transmission. \newpart{We assume that all the MAC layer FIFO queues are initially empty.} We restrict our scope to  stationary policies.

Let $\pi$ be a stationary policy with link activation frequency $f_e$ for link $e$. Then the service of the link $e$'s MAC layer FIFO queue is Bernoulli at rate $\gamma_e f_e$. 
The buffered source (link $e$), in effect, behaves as a discrete time FIFO G/Ber/1 queue. In the following subsection, we derive peak and average age for a discrete time G/Ber/1 queue. We use these results for the network case in sub-sections~\ref{sec:buffered_ber} and~\ref{sec:buffered_D}.

\subsection{Discrete Time G/Ber/1 Queue}
\label{sec:GBer1}
Consider a discrete time FIFO queue with Bernoulli service at rate $\mu$. Let the source generate update packets at epoches of a renewal process. Let $X$ denote the inter-arrival time random variable with general distribution $F_{X}$. Note that $X$ takes values in $\{1, 2, \ldots \}$.  We assume $\lambda = \EX{X}^{-1} < \mu$.

We derive peak and average age for this G/Ber/1 queue. Age for continuous time FIFO M/M/1 and D/M/1 queues was analyzed in~\cite{2012Infocom_KaulYates}. We will see that the results for the discrete time FIFO Ber/Ber/1 and D/Ber/1, which can be obtained as special cases of our G/Ber/1 results, differ from their continuous time counterparts, namely M/M/1 and D/M/1.

Let $T$ denote the system time for a packet, namely, the time from the packet's arrival to the time it completes service. To help derive peak and average age, we first analyze the system time $T$ for a packet in the G/Ber/1 queue.
\begin{framed}
\begin{lemma}
\label{lem:system_time}
The system time, $T$, in a FIFO G/Ber/1 queue is geometrically distributed with rate $\alpha^{\ast}$, where $\alpha^{\ast}$ is the solution to the equation
\begin{equation}
\label{eq:lem:system_time}
\alpha = \mu - \mu M_{X}\left(\log (1-\alpha) \right),
\end{equation}
where $M_{X}\left( \alpha \right) = \EX{e^{\alpha X}}$ denotes the moment generating function of the inter-arrival time $X$.
\end{lemma}
\end{framed}
\begin{IEEEproof}
See Appendix~\ref{pf:lem:system_time}, in the supplementary material.
\end{IEEEproof}
Note that the $\alpha^{\ast}$ depends on the distribution $F_X$.
Using Lemma~\ref{lem:system_time}, we can now compute peak and average age for the G/Ber/1 queue.
\begin{framed}
\begin{theorem}
\label{thm:GGeo1}
For G/Ber/1 queue with update packet generation rate $\lambda$ and service rate $\mu$ the peak age is given by
\begin{equation}
\label{eq:thm:GGeo1_peak}
A^{\text{p}} = \frac{1}{\alpha^{\ast}} + \frac{1}{\lambda},
\end{equation}
while the average age is given by
\begin{equation}\label{eq:thm:GGeo1_ave}
A^{\text{ave}} = \lambda\left[ \frac{M^{''}_{X}(0)}{2} + \frac{1}{\alpha^{\ast}}M_{X}^{'}\left( \log(1-\alpha^{\ast}) \right)\right] + \frac{1}{\mu} + \frac{1}{2},
\end{equation}
where $\alpha^{\ast}$ is given by~\eqref{eq:lem:system_time}, and $M_{X}(\alpha) = \EX{e^{\alpha X}}$ is the moment generating function of the inter-generation time $X$.
\end{theorem}
\end{framed}
\begin{IEEEproof}
\newpart{The peak age is given by
\begin{equation}
A^{\text{p}} = \EX{ T + X },
\end{equation}
where $T$ is the steady state system time, and $X$ is the inter-arrival time of update packets. This follows from the same line of arguments as in~\cite{2015ISIT_LongBoEM, 2014ISIT_CostaEp}, which derive the peak age for continuous time queues.} Since $\EX{X} = \frac{1}{\lambda}$ and $\EX{T} = \frac{1}{\alpha^{\ast}}$, form Lemma~\ref{lem:system_time}, the result follows. The proof for $A^{\text{ave}}$ is given in Appendix~\ref{pf:thm:GGeo1}, in the supplementary material.
\end{IEEEproof}

\begin{figure}
  \centering
  \includegraphics[width=0.95\linewidth]{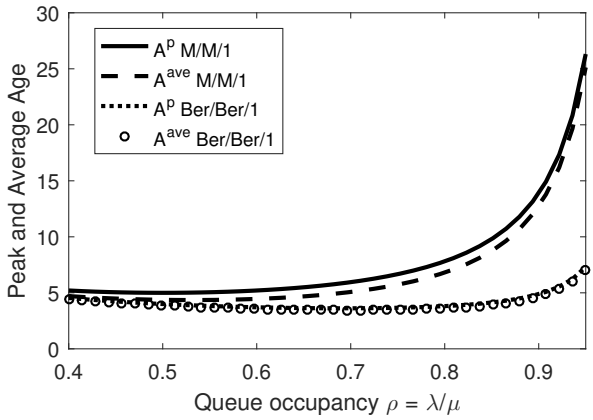}
  \caption{Plot of peak and average age as a function of queue occupancy $\rho = \lambda/\mu$ for various continuous time and discrete time queues. $\mu = 0.8$.}
  \label{fig:AgeComp_DTvsCT}
\end{figure}

The peak and average age for the continuous time M/M/1 queue was derived \newpart{in~\cite{2012Infocom_KaulYates, 2014ISIT_CostaEp, 2015ISIT_LongBoEM}}.
\newpart{We now show that the average/peak age for the discrete time queue differs significantly from its continuous time counterpart.}
In Figure~\ref{fig:AgeComp_DTvsCT}, we plot the peak and average age for the continuous time M/M/1 and the discrete time Ber/Ber/1 queue, as a function of queue occupancy $\rho = \lambda/\mu$. \newpart{Here, the service rate $\mu$ is set to $0.8$ for the sake of illustration only.}

We observe that the age difference between the continuous time and the discrete time cases can be very large. This is especially the case when the server utilization $\rho$ is close to $1$. We observe similar behavior when comparing the continuous time D/M/1 queue, with the discrete time D/Ber/1 queue.

In the following, we consider two specific discrete time queues, namely, Ber/Ber/1 and D/Ber/1. We show that the peak and average age, for these discrete time queues, can be upper-bounded by their continuous time counterparts M/M/1 and D/M/1. Although, this bound can be very weak (see Figure~\ref{fig:AgeComp_DTvsCT}), the optimal queue occupancy that minimizes the upper-bound yields a nearly optimal performance. We will use this, to then obtain a near optimal rate control and scheduling policy for the buffered source case.

\subsection{Bernoulli Generation of Update Packets}
\label{sec:buffered_ber}
Using Theorem~\ref{thm:GGeo1}, we now derive peak and average age for Bernoulli packet generation. Let $\lambda_e$ be the packet generation rate for link $e$. If link $e$ is getting served at link activation frequency $f_e$ under a stationary policy $\pi$, then its peak age is given by
\begin{equation}
\label{eq:Ber_peak}
A^{\text{p}}_{e}(f_e, \rho_e) = \left\{\!\!\!\! \begin{array}{cc}
\frac{1}{\gamma_e f_e}\left[ \frac{1}{\rho_e} + \frac{1}{1-\rho_e}\right] - \frac{\rho_e}{1 - \rho_e}, &\!\!\!\!\!\text{if}~\gamma_e f_e < 1 \\
1 + \frac{1}{\rho_e}, &\!\!\!\!\!\!\!\text{if}~\gamma_e f_e = 1
\end{array}\right.,
\end{equation}
while its average age is given by
\begin{equation}
\label{eq:Ber_ave}
A^{\text{ave}}_{e}(f_e, \rho_e) \!= \!\left\{\!\!\!\! \begin{array}{cc}
\frac{1}{\gamma_e f_e}\left[ 1 + \frac{1}{\rho_e} + \frac{\rho^{2}_{e}}{1-\rho_{e}}\right]\! - \!\frac{\rho^{2}_{e}}{1-\rho_{e}}, &\!\!\!\!\!\text{if}~\gamma_e f_e < 1 \\
1 + \frac{1}{\rho_e}, &\!\!\!\!\!\!\!\text{if}~\gamma_e f_e = 1
\end{array}\right.,
\end{equation}
where $\rho_e = \frac{\lambda_e}{\gamma_e f_e}$. See Appendix~\ref{pf:cor:BerGeo1}, in the supplementary material, for a detailed derivation.

In order to minimize AoI, unlike in the active source case, we need to jointly optimize over packet generation rates $\lambda_e$, or $\rho_e$, and scheduling policy $\pi \in \Pi_{\text{st}}$. Note that we optimize over the space of all stationary scheduling policies. Using~\eqref{eq:Ber_peak}, the peak age minimization problem is given by
\begin{align}
\begin{aligned}
\label{eq:peak_opt_buffered}
A^{\text{p}\ast}_{\mathcal{B}} =& \underset{\mathbf{f}, \bm{\rho} \in [0,1]^{|E|}}{\text{Minimize}}
& & \sum_{e \in E} w_{e} A^{\text{p}}_{e}(f_{e}, \rho_e) \\
& \text{subject to}
& & \mathbf{f} \in \mathcal{F}
\end{aligned}
\end{align}
where $\mathcal{F}$ is the set of feasible link activation frequencies; see~\eqref{eq:mathcalF}. Similarly, the average age minimization problem is given by
\begin{align}
\begin{aligned}
A^{\text{ave}\ast}_{\mathcal{B}} =& \underset{\mathbf{f}, \mathbf{\rho}\in [0,1]^{|E|} }{\text{Minimize}}
& & \sum_{e \in E} w_{e} A^{\text{ave}}_{e}(f_{e}, \rho_e) \\
& \text{subject to}
& & \mathbf{f} \in \mathcal{F}
\end{aligned}
\end{align}
We now derive an important separation principle which leads to a simple and practical solution to these problems.

The peak and average age for link $e$ can be upper bounded by as follows:
\begin{equation}
\label{eq:BerGeo1_peak_ub}
A^{\text{p}}_{e}(f_e, \rho_e) \leq \frac{1}{\gamma_e f_e}\left[ \frac{1}{\rho_e} + \frac{1}{1-\rho_e}\right],
\end{equation}
and
\begin{equation}
\label{eq:BerGeo1_ave_ub}
A^{\text{ave}}_{e}(f_e, \rho_e) \leq \frac{1}{\gamma_e f_{e}}\left[ 1 + \frac{1}{\rho_{e}} + \frac{\rho^{2}_{e}}{1-\rho_{e}} \right].
\end{equation}
The upper bounds in~\eqref{eq:BerGeo1_peak_ub} and~\eqref{eq:BerGeo1_ave_ub} are, in fact, the peak age and average age for the M/M/1 queue~\cite{2012Infocom_KaulYates, 2015ISIT_LongBoEM}. We define $\bar{\rho}^{\text{p}}$ and $\bar{\rho}^{\text{ave}}$ to be the optimal $\rho_e$ that minimizes the peak and average age upper-bounds, respectively. We have
\begin{equation}
\bar{\rho}^{\text{p}} = \arg\!\!\!\!\min_{\rho_e \in (0,1)}~~\frac{1}{\rho_e} + \frac{1}{1-\rho_e},
\end{equation}
and
\begin{equation}
\bar{\rho}^{\text{ave}} = \arg\!\!\!\!\min_{\rho_e \in (0,1)}~~1 + \frac{1}{\rho_{e}} + \frac{\rho^{2}_{e}}{1-\rho_{e}}.
\end{equation}
It is easy to see that $\bar{\rho}^{\text{p}} = \frac{1}{2}$, whereas $\bar{\rho}^{\text{ave}}$, is known to solve the equation $\rho^4 - 2\rho^3 + \rho^2 - 2\rho + 1 = 0$, and is approximately given by $\rho \approx 0.53$~\cite{2012Infocom_KaulYates}. The significance of $\bar{\rho}^{\text{p}}$ and $\bar{\rho}^{\text{ave}}$ is due to the following lemma.

\begin{framed}
\begin{lemma}
\label{lem:mm1_bound}
The peak and average age at $\rho_e = \bar{\rho}^{\text{p}}$ and $\rho_e = \bar{\rho}^{\text{ave}}$, respectively, is at most a unit away from the optimal:
\begin{equation}
\label{eq:nx1}
A^{\text{p}}_{e}\left(f_e, \bar{\rho}^{\text{p}}\right) - \min_{\rho \in [0,1]} A^{\text{p}}_{e}\left(f_e, \rho\right) \leq 1,
\end{equation}
and
\begin{equation}
\label{eq:nx2}
A^{\text{ave}}_{e}\left(f_e, \bar{\rho}^{\text{ave}}\right) - \min_{\rho \in [0,1]}A^{\text{ave}}_{e}\left(f_e, \rho\right) \leq 1,
\end{equation}
for all $f_e \in (0, 1)$.
\end{lemma}
\end{framed}
\begin{IEEEproof}
See Appendix~\ref{pf:lem:mm1_bound}, in the supplementary material.
\end{IEEEproof}

Motivated by this we propose the following separation principle policy (SPP):
\begin{framed}
\textbf{Separation principle policy:}
\begin{enumerate}
  \item Schedule links according to the stationary policy $\pi_C$ that minimizes peak age in the active source case. Here, $\pi_C$ is obtained as a solution to problem~\eqref{eq:opt_peak_age_problem}.
  \item Choose $\rho_e = \overline{\rho}$, for all $e \in E$.
  \item Generate update packets according to a Bernoulli process of rate $\lambda_e = \overline{\rho}\gamma_e f_e$.
\end{enumerate}
\end{framed}
Note that the rate control $\rho_e = \overline{\rho}$ is the same for all links. We choose $\bar{\rho} = \bar{\rho}^{\text{p}}$ to minimize peak age and $\bar{\rho} = \bar{\rho}^{\text{ave}}$ to minimize the average age.
We now prove that the SPP is close to the optimal peak and average age, namely, $A^{\text{p}\ast}_{\mathcal{B}}$ and $A^{\text{ave}\ast}_{\mathcal{B}}$, respectively.
\begin{framed}
\begin{theorem}
\label{thm:Q-IID}
Let $\mathbf{f}^{\ast}$ be the link activation frequency vector of the stationary policy $\pi_C$.
\begin{enumerate}
  \item Peak age of the stationary policy $\pi_C$ with rate control $\rho_e = \bar{\rho}^{\text{p}}$ is bounded by
      \begin{equation}
      A^{\text{p}}(\mathbf{f}^{\ast}, \bar{\rho}^{\text{p}}\mathbf{1}) \leq A^{\text{p}\ast}_{\mathcal{B}} + 1.
      \end{equation}

  \item Average age of the stationary policy $\pi_C$ with rate control $\rho_e = \bar{\rho}^{\text{ave}}$ is bounded by
      \begin{equation}
      A^{\text{ave}}(\mathbf{f}^{\ast}, \bar{\rho}^{\text{ave}}\mathbf{1}) \leq A^{\text{ave}\ast}_{\mathcal{B}} + 1.
      \end{equation}
\end{enumerate}
\end{theorem}
\end{framed}
\begin{IEEEproof}
See Appendix~\ref{pf:thm:Q-IID}, in the supplementary material.
\end{IEEEproof}

Theorem~\ref{thm:Q-IID}, therefore, says that when we restrict to stationary policies, separation between rate control and scheduling is nearly optimal. That is, if the rate control (choosing $\rho_e$) is performed assuming that there are no other contending links, and link scheduling is done by assuming active sources then the resulting solution is close to optimal.

\subsection{Periodic Generation of Update Packets}
\label{sec:buffered_D}
Since the packet generation is entirely a design parameter, we now consider the case where the sources generate update packets periodically. We will derive optimal packet generation period $D_e$, or equivalently rate $\lambda_e = 1/D_e$, for each link $e$ along with a scheduling policy in order to minimize age.

Using Theorem~\ref{thm:GGeo1} we can obtain peak and average age for a link $e$ under any stationary policy $\pi$ and periodic arrivals. Let $f_e$ be the link activation frequency for link $e$ under policy $\pi$, and $D_e$ be the period of packet generation for link $e$. Then the peak age is given by
\begin{equation}
\label{eq:D_peak}
A^{\text{p}}\left(f_e, \rho_e\right) = \frac{1}{\gamma_e f_e}\left[ \frac{1}{\rho_e} + \frac{1}{\sigma^{\ast}_{e}}\right],
\end{equation}
where $\rho_e = \left( D_e \gamma_e f_e \right)^{-1}$ and $\sigma^{\ast}_{e}$ is the solution to the equation $\sigma = 1 - \left(1 - \sigma\gamma_{e}f_{e} \right)^{D_{e}}$. Note that the new variable $\sigma^{\ast}_e$ is nothing but $\alpha^{\ast}/(\gamma_e f_e)$, for the $\alpha^{\ast}$ in Theorem~\ref{thm:GGeo1}. Average age for the same link $e$ is given by
\begin{equation}
\label{eq:D_ave}
A^{\text{ave}}_{e}\left(f_e, \rho_e\right) = \frac{1}{\gamma_e f_e}\left[ \frac{1}{2\rho_e} + \frac{1}{\sigma^{\ast}_{e}}\right] + \frac{1}{2}.
\end{equation}
See Appendix~\ref{pf:cor:DGeo1} for a detailed derivation.

Our objective is to minimize $\sum_{e \in E} w_e A^{\text{p}}_{e}(f_e, \rho_e)$ for peak age and $\sum_{e \in E} w_e A^{\text{ave}}_{e}(f_e, \rho_e)$ for average age, where $\mathbf{f}$ and $\bm{\rho}$ take values in $\mathcal{F}$ and $[0,1]^{|E|}$, respectively. Let $A^{\text{p}\ast}_{\mathcal{D}}$ and $A^{\text{ave}\ast}_{\mathcal{D}}$ denote the minimum peak and average age, jointly optimized over rate control $\mathbf{\rho}$ and stationary scheduling policies $\pi \in \Pi_{\text{st}}$. We first upper-bound peak and average age just as we did for the Bernoulli packet generation case.
\begin{framed}
\begin{lemma}
\label{lem:DGeo1_age_ub}
The peak and average age for link $e$ is upper bounded by
\begin{equation}
\label{eq:ub_peak_D}
A^{\text{p}}_{e}\left(f_e, \rho_e\right) \leq \frac{1}{\gamma_e f_e}\left[ \frac{1}{\rho_{e}} + \frac{1}{\hat{\sigma}_{e}} \right],
\end{equation}
and
\begin{equation}
\label{eq:ub_ave_D}
A^{\text{ave}}_{e}\left(f_{e}, \rho_{e}\right) \leq \frac{1}{\gamma_e f_e}\left[ \frac{1}{2\rho_e} + \frac{1}{\hat{\sigma}_{e}}\right] + \frac{1}{2},
\end{equation}
where $\hat{\sigma}_{e}$ solves $\hat{\sigma}_{e} = 1 - e^{-\frac{\hat{\sigma}_{e}}{\rho_{e}}}$.
\end{lemma}
\end{framed}
\begin{IEEEproof}
Using the fact that $1-x \leq e^{-x}$, we have
\begin{equation}
1 - (1 - f_e\gamma_e \sigma)^{D_e} \geq 1 - e^{-\sigma/\rho_e},
\end{equation}
where $\rho_e = \frac{1}{D_e \gamma_e f_e}$. This implies that if $\sigma^{\ast}$ solves $\sigma = 1 - (1 - \gamma_e f_e \sigma)^{D_e}$ and $\hat{\sigma}$ solves $\sigma = 1 - e^{-\sigma/\rho_e}$ then $\hat{\sigma} \leq \sigma^{\ast}$. The result follows from this.
\end{IEEEproof}

It is important to note that $\sigma^{\ast}$ in~\eqref{eq:D_peak} and~\eqref{eq:D_ave} is different from $\hat{\sigma}$ in Lemma~\ref{lem:DGeo1_age_ub}. In particular, $\sigma^{\ast}$ depends on $f_e$ and $D_e$ while $\hat{\sigma}$ is a function only of the queue occupancy $\rho_e = \frac{1}{D_e \gamma_e f_e}$. As a result the $\rho_e$ that minimizes the upper-bound(s), in Lemma~\ref{lem:DGeo1_age_ub}, is independent of $f_e$.

Furthermore, the upper-bounds in Lemma~\ref{lem:DGeo1_age_ub} are nothing but the peak and average age expressions for the continuous time D/M/1 queue~\cite{2012Infocom_KaulYates, talak18_determinacy, Inoue17_FCFS_AoIDist}. Let $\overline{\rho}^{\text{p}}$ and $\overline{\rho}^{\text{ave}}$  minimizes the peak age and average age upper-bounds in~\eqref{eq:ub_peak_D} and~\eqref{eq:ub_ave_D}, respectively. Numerically, it can be observed that $\overline{\rho}^{\text{p}} \approx 0.594$ and $\overline{\rho}^{\text{ave}} \approx 0.515$~\cite{2012Infocom_KaulYates}. We now make the following observation.
\begin{framed}
\begin{result}
\label{res:DGeo1_peak_gap}
The peak and average age at $\rho_e = \bar{\rho}^{\text{p}}$ and $\rho_e = \bar{\rho}^{\text{ave}}$, respectively, is at most a unit away from the optimal:
\begin{equation}
A^{\text{p}}_{e}\left(f_e, \overline{\rho}^{\text{p}}\right) - \min_{\rho \in [0,1]} A^{\text{p}}_{e}\left(f_{e}, \rho\right) \leq 1,
\end{equation}
and
\begin{equation}
A^{\text{ave}}_{e}\left(f_e, \overline{\rho}^{\text{ave}}\right) - \min_{\rho \in [0, 1]} A^{\text{ave}}_{e}\left(f_e, \rho\right) \leq 1,
\end{equation}
for all $f_e \in (0, 1)$.
\end{result}
\end{framed}
\begin{IEEEproof}
We note that the age difference:
\begin{equation}
\Delta^{\text{p}}(\gamma_e f_e) = A^{\text{p}}_{e}(f_e, \overline{\rho}^{\text{p}}) - \min_{\rho_e} A^{\text{p}}_{e}(f_e, \rho_e), \nonumber
\end{equation}
is a single variable function of the link $e$'s successful activation frequency $\gamma_ef_e$. Furthermore, $\gamma_ef_e$ can take values only in $(0, 1)$.
\begin{figure}
  \centering
  \includegraphics[width=0.95\linewidth]{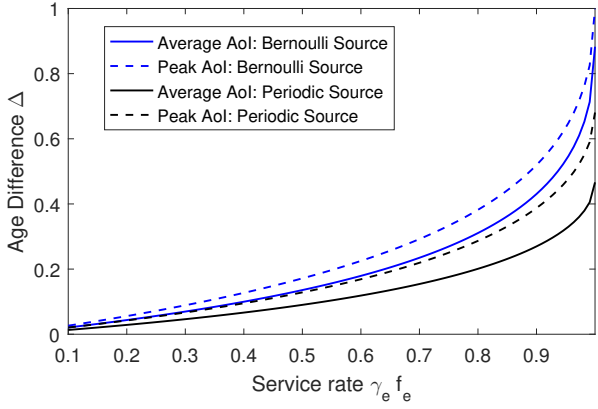}
  \caption{Plot of age difference $\Delta = A_{e}(f_{e}, \overline{\rho}) - \min_{\rho_e}A_{e}(f_{e}, \rho_e)$ as a function of service rate $\gamma_e f_e$ for Bernoulli and periodic packet generation.}\label{fig:PlotAgeBoundSim}
\end{figure}
In Figure~\ref{fig:PlotAgeBoundSim}, we plot this age difference, for both peak and average age as a function of $\gamma_e f_e$. We see that the age difference $\Delta$ is always below $1$, thereby, validating our result. In fact, $\Delta$ for peak age is observed to be less than $0.7$ and $\Delta$ for average age is observed to be less than $0.6$.
\end{IEEEproof}

Motivated by this we again resort to the same SPP as in Section~\ref{sec:buffered_ber} but with different rate control $\rho_e$ which \newpart{minimize} upper bounds in~\eqref{eq:ub_peak_D} and~\eqref{eq:ub_ave_D}. We prove the following bounds for this SPP.
\begin{framed}
\begin{result}
\label{thm:QD-IID}
Let $\mathbf{f}^{\ast}$ be the link activation frequency vector of the stationary policy $\pi_C$.
\begin{enumerate}
  \item Peak age of the stationary policy $\pi_C$ with rate control $\rho_e = \overline{\rho}^{\text{p}}$ is bounded by
      \begin{equation}
      A^{\text{p}}(\mathbf{f}^{\ast}, \overline{\rho}^{\text{p}}\mathbf{1}) \leq A^{\text{p}\ast}_{\mathcal{D}} + 1.
      \end{equation}

  \item Average age of the stationary policy $\pi_C$ with rate control $\rho_e = \overline{\rho}^{\text{ave}}$ is bounded by
      \begin{equation}
      A^{\text{ave}}(\mathbf{f}^{\ast}, \overline{\rho}^{\text{ave}}\mathbf{1}) \leq A^{\text{ave}\ast}_{\mathcal{D}} + 1.
      \end{equation}
\end{enumerate}
\end{result}
\end{framed}
\begin{IEEEproof}
Using Result~\ref{res:DGeo1_peak_gap}, the proof follows the same line of arguments as that of Theorem~\ref{thm:Q-IID}.
\end{IEEEproof}

\section{Performance Bounds for SPP}
\label{sec:non-stationary}
In Section~\ref{sec:queued}, we considered two update generation models, namely Bernoulli and periodic. We derived the SPP, that optimizes age jointly over update generation rate and scheduling policy. However, we limited scheduling policies to the space of stationary scheduling policies $\Pi_{\text{st}}$. We now consider a much larger space of scheduling policies.

It is conceivable that a policy that schedules links depending on the link's age or buffer backlogs, may perform significantly better then the optimal stationary policies. Let $Q_e(t)$ denote the queue length of the buffer of source $e$, and $\mathbf{Q}(t) = \left( Q_{e}(t)\right)_{e \in E}$. Define history $\mathcal{H}_{Q}(t)$ to be
\begin{equation}\label{eq:history_q}
\mathcal{H}_{Q}(t) = \mathcal{H}(t)\bigcup\{\mathbf{Q}(\tau')~|~0\leq \tau' \leq t \},
\end{equation}
where $\mathcal{H}(t)$ is as defined in~\eqref{eq:history}. Let $\Pi_Q$ denote the space of all scheduling policies which base its scheduling decision at time $t$ on the history $\mathcal{H}_{Q}(t)$, for all $t$.

We now consider the age of the SPP, with the minimum age achieved by jointly optimizing over the update generation rate control and scheduling policy $\pi \in \Pi_{Q}$. We first consider Bernoulli update generation and then periodic update generation.

\subsection{Bernoulli Generation of Update Packets}
Let each source generate update packets according to a Bernoulli process. Let $A^{\text{p}\ast}_{\mathcal{QB}}$ and $A^{\text{ave}\ast}_{\mathcal{QB}}$ denote the optimal peak and average age, achieved by jointly optimizing over the update generation rates and the scheduling policy $\pi \in \Pi_Q$.
We now show that the proposed separation principle policy is at most a constant factor away from this optimal age.
\begin{framed}
\begin{corollary}\label{cor:Q-IID}
Let $\mathbf{f}^{\ast}$ be a vector of link activation frequencies under the stationary policy $\pi_C$.
\begin{enumerate}
  \item Peak age of the stationary policy $\pi_C$ with rate control $\rho_e = \bar{\rho}^{\text{p}}$ is bounded by
      \begin{equation}
      A^{\text{p}}\left( \mathbf{f}^{\ast}, \bar{\rho}^{\text{p}}\mathbf{1} \right) \leq \left( \frac{1}{\bar{\rho}^{\text{p}}} + \frac{1}{1 - \bar{\rho}^{\text{p}} }\right) A^{\text{p}\ast}_{\mathcal{QB}}.
      \end{equation}

  \item Average age of the stationary policy $\pi_C$ with rate control $\rho_e = \bar{\rho}^{\text{ave}}$ is bounded by
      \begin{equation}
      A^{\text{ave}}\left( \mathbf{f}^{\ast}, \bar{\rho}^{\text{ave}}\mathbf{1} \right) \leq 2\left( 1 + \frac{1}{\bar{\rho}^{\text{ave}}} + \frac{(\bar{\rho}^{\text{ave}})^2}{1 - \bar{\rho}^{\text{ave}}} \right) A^{\text{ave}\ast}_{\mathcal{QB}}.
      \end{equation}
\end{enumerate}
\end{corollary}
\end{framed}
\begin{IEEEproof}
See Appendix~\ref{pf:cor:Q-IID}, in the supplementary material.
\end{IEEEproof}
We know that $\bar{\rho}^{\text{p}} = 1/2$ and $\bar{\rho}^{\text{ave}} \approx 0.53$. Thus, the peak age of the stationary policy $\pi_C$, with rate control $\rho_e = 0.5\gamma_e f_e$, is at most a factor of $4$ away from optimality. Also, the average age of the stationary policy $\pi_C$, with rate control $\lambda_e = \bar{\rho}^{\text{ave}} \gamma_e f_e$, is at most a factor of $2\left( 1 + \frac{1}{\bar{\rho}^{\text{ave}}} + \frac{(\bar{\rho}^{\text{ave}})^2}{1 - \bar{\rho}^{\text{ave}}} \right) \approx 7$ away from optimality.
It is important to note that the constant factors of optimality in Corollary~\ref{cor:Q-IID} are independent of the network size.

We next consider periodic update generation, which yield much smaller factors of optimality, than derived here for the Bernoulli update generation.

\subsection{Periodic Generation of Update Packets}
Let the update generation be periodic at each source. Let $A^{\text{p}\ast}_{\mathcal{QD}}$ and $A^{\text{ave}\ast}_{\mathcal{QD}}$ denote the minimum peak and average age that can be achieved by jointly optimizing over the update generation rate and the scheduling policy $\pi \in \Pi_Q$. We prove that our SPP is at most a constant factor away from this optimal age.
\begin{framed}
\begin{result}
Let $\mathbf{f}^{\ast}$ be the vector of link activation frequencies for policy $\pi_C$.
\begin{enumerate}
  \item Peak age of the stationary policy $\pi_C$ with rate control $\rho_e = \overline{\rho}^{\text{p}}$ is bounded by
      \begin{equation}
      A^{\text{p}}\left( \mathbf{f}^{\ast}, \overline{\rho}^{\text{p}}\mathbf{1}\right) \leq \left( \frac{1}{\hat{\sigma}} + \frac{1}{\overline{\rho}^{\text{p}}} \right) A^{\text{p}\ast}_{\mathcal{QD}}.
      \end{equation}

  \item Average age of the stationary policy $\pi_C$ with rate control $\rho_e = \overline{\rho}^{\text{ave}}$ is bounded by
      \begin{equation}
      A^{\text{ave}}\left( \mathbf{f}^{\ast}, \overline{\rho}^{\text{ave}}\mathbf{1}\right) \leq 2\left( \frac{1}{2\hat{\sigma}} + \frac{1}{\overline{\rho}^{\text{ave}}} \right) A^{\text{ave}\ast}_{\mathcal{QD}}.
      \end{equation}
\end{enumerate}
Here, $\hat{\sigma} \in (0,1)$ is the unique solution to $\sigma = 1 - e^{-\sigma/\overline{\rho}^{\text{p}}}$.
\end{result}
\end{framed}
\begin{IEEEproof}
Using Result~\ref{res:DGeo1_peak_gap} and~\ref{thm:QD-IID}, the proof follows the same line of arguments as that of Corollary~\ref{cor:Q-IID}.
\end{IEEEproof}

Computing the upper-bound factors numerically, we see that the peak age SPP policy is within a factor of $\approx 2.15$ from the optimal peak age, while the average age SPP policy is within a factor of $\approx 4.51$ from the optimal average age. In Table~\ref{tbl:bound} we summarize the performance of our SPP policies, under both Bernoulli and periodic packet generation, over the space of all policies. If the bounds are tight, then it suggests that periodic packet generation should perform much better than Bernoulli generation.
\begin{table}
\caption{Rate control for separation principle policy (SPP) $\overline{\rho} = \frac{\lambda_e}{f_e}$ and optimality of SPP over space of all policies.}
\label{tbl:bound}
\begin{center}
\begin{tabular}{|ccc|}
\hline
\textbf{Peak Age} & $\overline{\rho}$ & Factor of optimality \\
\hline
Bernoulli updates & $1/2$  & $4$\\
Periodic updates & $\approx 0.594$  & $\approx 2.15$ \\
\hline
\textbf{Average Age} &  Optimal~$\rho$ & Factor of optimality \\
\hline
Bernoulli updates & $\approx 0.53$ & $\approx 7$ \\
Periodic updates & $\approx 0.515$ & $\approx 4.51$ \\
\hline
\end{tabular}
\end{center}
\end{table}

\section{Numerical Results}
\label{sec:num}
We consider a $K$-link activation network with $N$ links. For this network, a feasible activation set $m$ contains at most $K$ links. A fraction $\theta$ of the links have bad channel with $\gamma_e = \gamma_{\text{bad}}$, while the rest have $\gamma_e = \gamma_{\text{good}} > \gamma_{\text{bad}}$. We let \newpart{$w_e = 1/N$} for all the links. Similar results are observed for single-hop interference network.

\subsection{Network with Active Sources}
First, consider the case in which all the sources in the network are active sources. We plot and compare the proposed peak age optimal policy $\pi_C$ (shown in red), a uniform stationary policy that schedules maximal subsets in $\mathcal{A}$ randomly with uniform probability, and a round robin policy (RR) that schedules $K$ links at a time.

\begin{figure}
  \centering
  \includegraphics[width=0.45\textwidth]{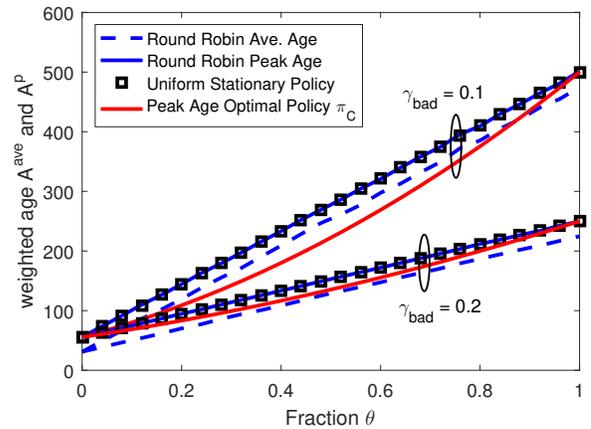}
  \caption{Plot of \newpart{weighted peak and average age} as a function of fraction of links, $\theta$, with bad channel. $N = 50$, $K = 1$, $\gamma_{\text{good}} = 0.9$, and $\gamma_{\text{bad}} = 0.1$ and $0.2$.} \label{fig:main3}
\end{figure}
Figure~\ref{fig:main3} considers the simplest case with $K = 1$, and \newpart{plots the weighted peak and average age, namely $A^{\text{p}}$ and $A^{\text{ave}}$,} as a function of $\theta$. Here, the network has $N = 50$ links, $\gamma_{\text{good}} = 0.9$, and two cases of $\gamma_{\text{bad}} = 0.1$ and $\gamma_{\text{bad}} = 0.2$ are plotted. Note that the peak age and average age coincide for stationary policies by Theorem~\ref{lem:rp}. We observe this in simulation. Thus, to reduce clutter, we have plotted only one curve for the peak age optimal policy $\pi_C$ and the uniform stationary policy.

We observe in Figure~\ref{fig:main3} that both peak and average age increases as $\theta$, the fraction of links with bad channel, increases. This is to be expected as with more error prone channels, it takes more time for the source to update the destination. For $\gamma_{\text{bad}} = 0.1$, we observe in Figure~\ref{fig:main3}, that the peak age optimal policy $\pi_C$ achieves the minimum peak age. Furthermore, when the channel statistics \newpart{is} more asymmetric, i.e. $\theta$ not near $0$ or $1$, the average age performance of the peak age optimal policy $\pi_C$ is better than the round robin and uniform stationary policy. We also observe that the round robin policy and uniform stationary policy achieve the same peak age. \newpart{This} validates Theorem~\ref{thm:peak_age_char} which states that any two policies with same link activation frequencies should have the same peak age.

In Figure~\ref{fig:main3}, we see that when the channel statistics across links is more symmetric (i.e., $\theta$ closer to $0$ or $1$), the round robin policy yields a slightly smaller average age than the peak age optimal policy $\pi_C$. In fact, when $\gamma_{\text{bad}}$ is increased to $0.2$, the round robin policy performs better in average age for all $\theta$. However, the average age optimal centralized scheduling policy is yet unknown even for this simple network (with $K = 1$), and hence by Theorem~\ref{thm:peak_ave}, the average age of the peak age optimal policy $\pi_C$ is at most factor $2$ away from the optimal average age, which is consistent with Figure~\ref{fig:main3}.

\begin{figure}
  \centering
  \includegraphics[width=0.45\textwidth]{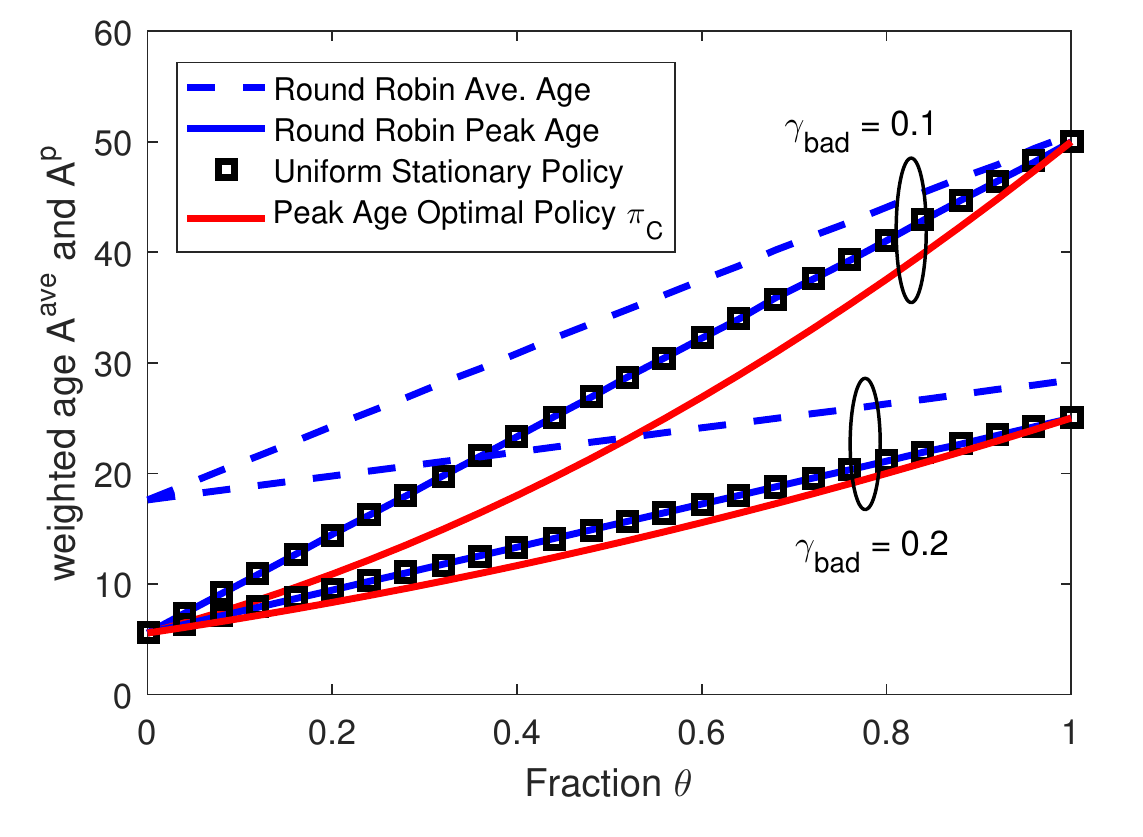}
  \caption{Plot of \newpart{weighted peak and average age} as a function of fraction of links, $\theta$, with bad channel. $N = 50$, $K = 10$, $\gamma_{\text{good}} = 0.9$, and $\gamma_{\text{bad}} = 0.1$ and $0.2$.} \label{fig:main2}
\end{figure}
This problem is exacerbated when we move to $K > 1$, in which case it is difficult to intuit a `good' policy that minimizes average age. Figure~\ref{fig:main2} \newpart{plots the weighted peak and average age} as a function of $\theta$. \newpart{Also, plotted is a round robin policy of period $T = \lceil N/K \rceil$, which schedules the $K$ worst channels in the first slot, the next $K$ worst channels in the second slot, and so on.
All the parameters are same as in Figure~\ref{fig:main3}, except that we can activate $K = 10$ links at a time. We observe that the proposed policy $\pi_C$ ensures peak age optimality, and also outperforms other simple scheduling policies in terms of its average age.} This observation is not limited to the policies presented here, but in general, as it is difficult to come up with average age optimal policies for a network with general interference constraints.

\subsection{Network with Buffered Sources}
\begin{figure}
  \centering
  \includegraphics[width=0.45\textwidth]{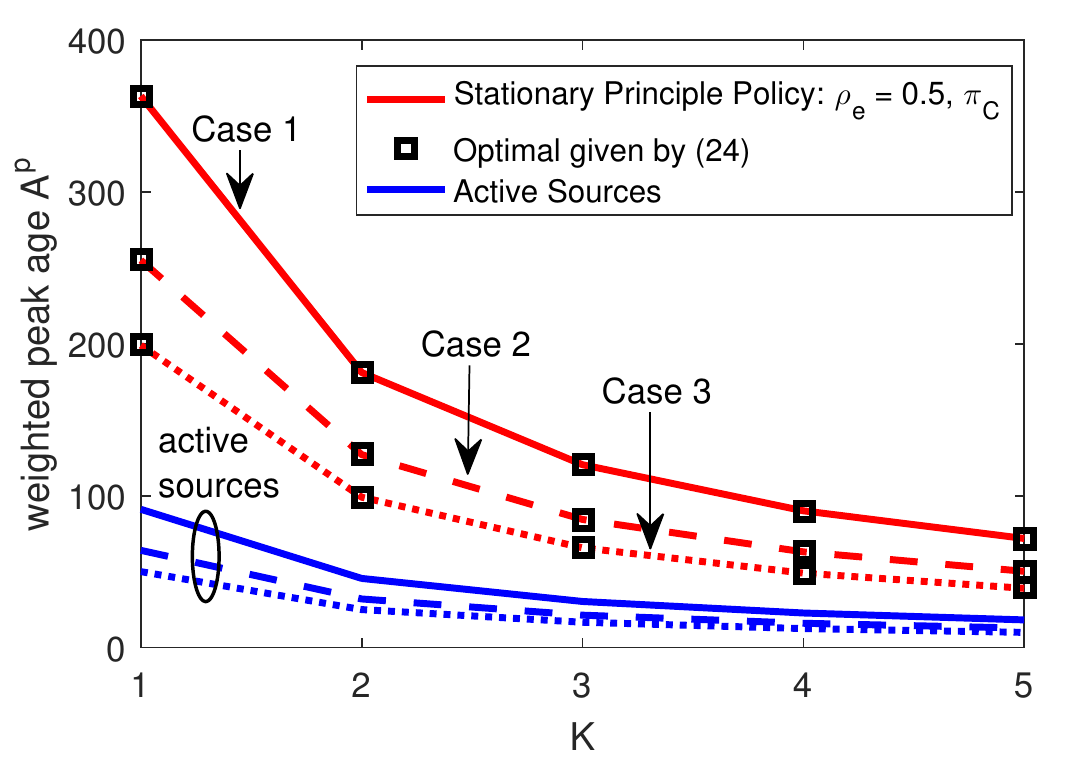}
  \caption{Plot of the weighted peak age $A^{\text{p}}$, for a network with buffered nodes, as a function of $K$. Case 1: $N = 50$, $\gamma_{\text{good}} = 0.9$, $\gamma_{\text{bad}} = 0.1$, $n_{\text{bad}} = 7$. Case 2: $N = 10$, $\gamma_{\text{good}} = 0.9$, $\gamma_{\text{bad}} = 0.1$, $n_{\text{bad}} = 7$. Case 3: $N = 50$ and $\gamma_e = 1$ for all links.} \label{fig:main6}
\end{figure}
We next consider the sources in the network to be buffered sources. We assume Bernoulli arrival of update packets. We plot three cases to illustrate the near optimality of the separation principle policy (SPP): In Case~1, we have $N=50$, $n_{\text{bad}} = 7$ links have bad channel, i.e., $\gamma_e = \gamma_{\text{bad}} = 0.1$ while the remaining have good channel $\gamma_e = \gamma_{\text{good}} = 0.9$. In Case~2, we have $N=10$, $n_{\text{bad}} = 7$, $\gamma_{\text{bad}} = 0.1$, and $\gamma_{\text{good}} = 0.9$. In Case~3, we consider $N = 50$ and $\gamma_e = 1$ for all links $e$. We let link weights to be unity, i.e. $w_e = 1$ for all $e$.

We compare the peak age SPP, which chooses $\rho_e = 1/2$ for every link and the link activation frequency $\mathbf{f}^{\ast}$ that solves~\eqref{eq:opt_klink}. In Figure~\ref{fig:main6}, we plot the peak age achieved by the peak age SPP and the optimal $A^{\text{p}\ast}_{\mathcal{B}}$ of~\eqref{eq:peak_opt_buffered}, obtained numerically.  We observe that the SPP nearly attains the optimal peak age for buffered node in~\eqref{eq:peak_opt_buffered} in all three cases.

This can be seen from our observation in Figure~\ref{fig:PlotAgeBoundSim}. In Figure~\ref{fig:PlotAgeBoundSim}, we observe that the age difference between optimal age and the age with rate control $\rho_e = 1/2$, which is $\Delta$, diminishes drastically as link activation frequency decreases. When the interference in the network is large, the link activation frequencies are bound to be small. This essentially results in close proximity of our separation principle policy with the optimal.

In Figure~\ref{fig:main6}, we also plot (in blue) peak age if the network had active sources instead of buffered sources. We observe that optimal peak age for the buffered case is about $4$ times that in the active source case. This shows that the cost of not being able to control the MAC layer queue can be as large as a $4$ fold increase in age.

\section{Conclusion}
\label{sec:conclusion}
We considered the problem of minimizing age of information in wireless networks, consisting a several source-destination communication links, under general interference constraints. For a network with active sources, where fresh updates are available for every transmission, we show that a stationary policy is peak age optimal, and is also within a factor of two of the optimal average age. For a network with buffered sources, in which the generated update packets are queued at the MAC layer queue for transmission, we proved a separation principle wherein it suffices to design scheduling and rate control separately. Numerical evaluation suggest that this proposed separation principle policy is nearly indistinguishable from the optimal. We also derived peak age and average age for discrete time FIFO G/Ber/1 queue, which may be of independent interest.

\bibliographystyle{ieeetr}

\appendix

\subsection{Proof of Theorem~\ref{thm:peak_age_char}}
\label{pf:thm:peak_age_char}
Let $\pi$ be a policy in $\Pi$, and $T_{e}(i)$ be the time of $i$th successful activation for link $e$. Then $S_{e}(i) = T_{e}(i) - T_{e}(i-1)$, for all $i \geq 1$, is the inter-(successful) activation time for link $e$, where $T_{e}(0) = 0$. Note that $S_{e}(i) = A_{e}(T_{e}(i))$ for all age update instances $i$. This implies that the peak age is given by
\begin{align}
\limsup_{N \rightarrow \infty} \frac{1}{N}\sum_{i=1}^{N} A_{e}\left( T_{e}(i)\right) &= \limsup_{N \rightarrow \infty} \frac{1}{N}\sum_{i=1}^{N} S_{e}(i), \\
&= \limsup_{N \rightarrow \infty} \frac{T_{e}(N)}{N}. \label{eq:rajat}
\end{align}
Notice that the time $T_{e}(N) \rightarrow \infty$ as $N \rightarrow \infty$. We, therefore, have
\begin{align}
\frac{1}{A_{e}^{\text{p}}(\pi)} = \liminf_{N \rightarrow \infty} \frac{N}{T_{e}(N)} &= \liminf_{T \rightarrow \infty} \frac{1}{T}\sum_{t=1}^{T} U_{e}(t)R_{e}(t), \label{eq:rajat2}
\end{align}
where $U_{e}(t) = \mathbb{I}_{\left\{e \in m(t), m(t) \in \mathcal{A} \right\}}$ and $R_{e}(t)$ is the channel process. Now, notice that the process $\left(U_{e}(t), R_{e}(t) \right)_{t \geq 0}$ is jointly ergodic. To see this, note that $\{U_{e}(t)\}_{t \geq 0}$ is an ergodic processes because $\{U_{e}(t)\}_{t \geq 0}$ takes values only in $\{0, 1\}$ and the limit in~\eqref{eq:link_act_freq} exists for $\pi \in \Pi$. Moreover, since $R_{e}(t)$ is independent of $U_{e}(t)$, they are jointly ergodic. We, therefore, have
\begin{align}
\liminf_{T \rightarrow \infty} \frac{1}{T}\sum_{t=1}^{T} U_{e}(t)R_{e}(t) &= \EX{\liminf_{T \rightarrow \infty} \frac{1}{T}\sum_{t=1}^{T} U_{e}(t)R_{e}(t)}, \nonumber \\
&= \liminf_{T \rightarrow \infty} \frac{1}{T}\sum_{t=1}^{T} \EX{U_{e}(t)R_{e}(t)}, \\
&= \liminf_{T \rightarrow \infty}\gamma_e \EX{\frac{1}{T}\sum_{t=1}^{T} U_{e}(t)}, \\
&= \gamma_e f_e(\pi),
\end{align}
where the first equality follows due to ergodicity, the second due to the bounded convergence theorem~\cite{Durrett}, and the third because $R_{e}(t)$ is i.i.d. across time $t$ with $\gamma_e = \EX{R_{e}(t)}$ and is independent of $U_{e}(t)$. The last equality follows from~\eqref{eq:link_act_freq}.
Weighted summation over all links $e \in E$ gives the result.

To prove that the peak age $A^{\text{p}}(\pi)$ can be achieved by a stationary centralized policy $\pi_{\text{st}} \in \Pi$, it suffices to show that a stationary centralized policy $\pi_{\text{st}}$ achieve the same link activation frequencies, i.e., $\mathbf{f}(\pi) = \mathbf{f}(\pi_{\text{st}})$.

Let $\pi \in \Pi$ be the policy that achieves link activation frequencies $\mathbf{f} = \left( f_{e} | e \in E\right)$. Then, the policy $\pi$ activates interference-free sets in $\mathcal{A}$ also with a certain frequency. Let $x_{m}$ be the frequency of activation for a set $m \in \mathcal{A}$, i.e.,
\begin{equation}
x_{m} = \limsup_{T \rightarrow \infty} \frac{1}{T}\sum_{t=1}^{T}\mathbb{I}_{\{m(t) = m\}},
\end{equation}
where $m(t)$ denotes the set of links activated at time $t$. Clearly, we should have
\begin{equation} \label{eq:z1}
\sum_{m \in \mathcal{A}} x_{m} \leq 1.
\end{equation}
Furthermore, we must have $\mathbf{f}(\pi) = M\mathbf{x}$, where $M$ is given by~\eqref{eq:M}, and $\mathbf{f}(\pi)$ and $\mathbf{x}$ are column vectors of $f_e(\pi)$ and $x_{m}$, respectively. Now consider a stationary centralized policy $\pi_{\text{st}} \in \Pi$ for which $m \in \mathcal{A}$ is activated in each slot with probability $x_{m}$, independent across slots; we can do this because of the property~\eqref{eq:z1}. Then we have $\mathbf{f}(\pi_{\text{st}}) = M\mathbf{x} = \mathbf{f}(\pi)$. This proves the result.

\subsection{Proof of Theorem~\ref{thm:age_relation}}
\label{pf:thm:age_relation}
The proof is a direct consequence of the Cauchy-Schwartz inequality. Consider a policy $\pi \in \Pi$ and let $T_{e}(i)$ be the time of $i$th successful activation for link $e$. Then $S_{e}(i) = T_{e}(i) - T_{e}(i-1)$, for all $i \geq 1$, is the inter-(successful) activation time for link $e$, where $T_{e}(0) = 0$. Note that $S_{e}(i) = A_{e}(T_{e}(i))$ for all age update instances $i$. This implies that the peak age is given by
\begin{align}
\limsup_{N \rightarrow \infty} \frac{1}{N}\sum_{i=1}^{N} A_{e}\left( T_{e}(i)\right) &= \limsup_{N \rightarrow \infty} \frac{1}{N}\sum_{i=1}^{N} S_{e}(i).
\label{eq:peak_age_expr}
\end{align}
The average is given by
\begin{align}
A^{\text{ave}}_{e}(\pi) &= \lim_{N \rightarrow \infty} \frac{\sum_{i=1}^{N} \sum_{k = 1}^{S_{e}(i)} k }{\sum_{i=1}^{N} S_{e}(i)} = \lim_{N \rightarrow \infty} \frac{\frac{1}{2}\sum_{i=1}^{N} S_{e}(i)^2}{\sum_{i=1}^{N} S_{e}(i)} + \frac{1}{2}. \label{eq:average_age_expr}
\end{align}
Cauchy-Schwartz inequality gives us 
\begin{equation}
\left( \sum_{i=1}^{N} S_{e}(i) \right)^{2} \leq N \sum_{i=1}^{N} S^{2}(i).
\end{equation}
Therefore, we must have
\begin{equation}
\frac{1}{2} \frac{1}{N} \sum_{i=1}^{N}S_{e}(i) \leq \frac{0.5 \sum_{i=1}^{N} S_{e}(i)^{2}}{\sum_{i=1}^{N} S_{e}(i)}.
\end{equation}
This with~\eqref{eq:peak_age_expr} and~\eqref{eq:average_age_expr} yields $\frac{1}{2}A^{\text{p}}_{e}(\pi) + \frac{1}{2} \leq A^{\text{ave}}_{e}(\pi)$. Note that we can claim this because $A^{\text{p}}(\pi)$ is finite for $\pi \in \Pi$ due to~\eqref{eq:local1}. Weighted summation over $e \in E$ gives the desired result.

\subsection{Proof of Lemma~\ref{lem:rp}}
\label{pf:lem:rp}
For a stationary policy, let $p$ be the probability that link $e$ is successfully activated in a time slot, i.e.,
\begin{equation}
p = \prob{e \in m(t),~m(t) \in \mathcal{A}},
\end{equation}
where $m(t)$ is the set of links activated at time $t$.
Since the policy is stationary, the inter-(successful) activation times $S_{e}(i)$ would be independent and geometrically distributed with rate $1/p$ given by:
$\prob{S_{e}(i) = k} = p\left( 1 - p \right)^{k-1}$,
for all $k \in \{1, 2, \ldots \}$. For this distribution we know that
$\EX{S_{e}(1)} = \frac{1}{p}$ and $\EX{S_{e}^{2}(1)} = \frac{2 - p}{p^2}$.
Using~\eqref{eq:average_age_expr} we obtain the average age of link $e$ to be
\begin{align}
\lim_{T \rightarrow \infty} \frac{1}{T}\sum_{t=1}^{T} A_{e}(t) &= \lim_{N \rightarrow \infty} \frac{\sum_{i=1}^{N} \frac{1}{2}S_{e}^{2}(i) }{\sum_{i=1}^{N} S_{e}(i)} +  \frac{1}{2}, \\
&= \frac{\frac{1}{2}\frac{2-p}{p^2}}{\frac{1}{p}} + \frac{1}{2} = \frac{1}{p}. \label{eq:y4}
\end{align}
Using~\eqref{eq:peak_age_expr} we obtain the peak age of the link $e$ to be
\begin{align}
\limsup_{N \rightarrow \infty} \frac{1}{N}\sum_{i=1}^{N} A_{e}\left( T_{e}(i)\right) &= \limsup_{N \rightarrow \infty} \frac{1}{N}\sum_{i=1}^{N} S_{e}(i), \\
&= \EX{S_{e}(1)} = \frac{1}{p}. \label{eq:y5}
\end{align}
The result can be obtained from~\eqref{eq:y4} and~\eqref{eq:y5} by weighted-averaging.


\subsection{Proof of Theorem~\ref{thm:opt_peak_age_char}}
\label{pf:thm:opt_peak_age_char}
Since dependence of $\gamma_e$ is only in the form $w_e/\gamma_e$, we assume $\gamma_e = 1$, for all $e$, for clarity of presentation. The dependence on $\gamma_e$ can be re-constructed by substituting $w_e/\gamma_e$ in place of $w_e$ in the following proof.

The peak age minimization problem~\eqref{eq:opt_peak_age_problem} can be re-written with the objective
\begin{equation}
\sum_{e \in E} \frac{w_e}{\sum_{m \in \mathcal{A}} M_{e,m} x_m},
\end{equation}
over variables $x_m$, for $m \in \mathcal{A}$, with constraints $\sum_{m \in \mathcal{A}} x_m \leq 1$ and $x_m \geq 0$ for all $m \in \mathcal{A}$. The Lagrangian function for this problem is
\begin{multline}
L(\mathbf{x}, \Omega, \bm{\nu}) = \sum_{e \in E} \frac{w_e}{\sum_{m \in \mathcal{A}} M_{e,m} x_m} + \Omega \left( \sum_{m \in \mathcal{A}} x_m - 1 \right) \\
+ \sum_{m \in \mathcal{A}} \nu_m x_m, \nonumber
\end{multline}
for $\Omega \geq 0$ and $\nu_m \geq 0$ for all $m \in \mathcal{A}$. The KKT conditions then imply
\begin{align}
&\frac{\partial L}{\partial x_m} = 0,~~\text{for all}~m \in \mathcal{A}, \label{eq:nt1}\\
&\Omega \left( \sum_{m \in \mathcal{A}} x_m - 1 \right) = 0, \label{eq:nt2} \\
&\nu_m x_m = 0~~\text{for all}~m \in \mathcal{A}, \label{eq:nt3}
\end{align}
along with feasibility constraints for $\mathbf{x}$, $\Omega \geq 0$, and $\nu_m \geq 0$ for all $m \in \mathcal{A}$. Now~\eqref{eq:nt1} implies
\begin{equation}
\frac{\partial L}{\partial x_m} = - \sum_{e \in E} \frac{w_e M_{e,m}}{\left( \sum_{m' \in \mathcal{A}} M_{e,m'} x_{m'}\right)^{2}} + \Omega - \nu_m = 0,
\end{equation}
which reduces to
\begin{equation}\label{eq:nt4}
\Omega_{m}(\mathbf{x}) = \Omega - \nu_m,
\end{equation}
for all $m \in \mathcal{A}$. Using~\eqref{eq:nt3} and~\eqref{eq:nt4} we get that if $x_m > 0$ then $\nu_m = 0$ which implies $\Omega_{m}(\mathbf{x}) = \Omega$, while $\Omega_{m}(\mathbf{x}) \leq \Omega$ for all $m \in \mathcal{A}$. This proves conditions~1 and~2 of Theorem~\ref{thm:opt_peak_age_char}.

Since $\mathbf{x}$ that satisfy the KKT conditions also solve~\eqref{eq:opt_peak_age_problem} we should have $f_e = \sum_{m \in \mathcal{A}} M_{e,m} x_m > 0$; otherwise the objective function would be unbounded. Thus, $\Omega_{m}(\mathbf{x}) = \Omega - \nu_m > 0$ which implies $\Omega > 0$ for all $m \in \mathcal{A}$. Then~\eqref{eq:nt2} implies $\sum_{m \in \mathcal{A}} x_m = 1$.

\newpart{We now prove that the $\Omega$ defined in Theorem~\ref{thm:opt_peak_age_char} is the optimal peak age $A^{\text{p}\ast}$.}
Given that $\mathbf{x}$ is the optimal solution to~\eqref{eq:opt_peak_age_problem} and $\Omega$ be as defined in Theorem~\ref{thm:opt_peak_age_char}, the optimal peak age is given by
\begin{align}
A^{\text{p}\ast} = \sum_{e \in E} \frac{w_e}{\left( M \mathbf{x}\right)_{e}} &= \sum_{e \in E} \frac{w_e}{\left( M \mathbf{x}\right)^{2}_{e}}\left(M\mathbf{x}\right)_{e},\\
&= \sum_{e \in E} \frac{w_e}{\left( M \mathbf{x}\right)^{2}_{e}} \sum_{m \in \mathcal{A}} M_{e,m}x_m.
\end{align}
Exchanging the two summations we get
\begin{align}
A^{\text{p}\ast} &= \sum_{m \in \mathcal{A}} x_{m} \sum_{e \in E} \frac{w_e M_{e,m}}{  \left(M \mathbf{x}\right)^{2}_{e} }, \\
&= \sum_{m \in \mathcal{A}, x_{m} > 0} x_{m} \sum_{e \in m} \frac{w_e}{  \left(M \mathbf{x}\right)^{2}_{e} },
\end{align}
where the last equality follows from the definition of $M_{e,m}$. Notice that $\sum_{e \in m} \frac{w_e}{  \left(M \mathbf{x}\right)^{2}_{e} }$ is in fact $\Omega_{m}(\mathbf{x})$, which equals $\Omega$ since $x_{m} > 0$. This gives
\begin{equation}
A^{\text{p}\ast} = \sum_{m \in \mathcal{A}, x_m > 0} x_m \Omega = \Omega,
\end{equation}
where the last equality follows because $\sum_{m \in \mathcal{A}}x_m = 1$.
\subsection{Derivation of Peak and Average Age for D/Ber/1 Queue}
\label{pf:cor:DGeo1}
Let $D$ be the period of the periodic packet generation and $\mu$ be the Bernoulli service rate. The probability distribution of inter-arrival time $X$ of update packets is given by $\pr{X = D} = 1$ and $\pr{X = k} = 0$ for all $k \neq D$. Thus, $M_{X}(t) = e^{Dt}$. Using this, the equation~\eqref{eq:lem:system_time} can be written as
\begin{align}
\alpha^{\ast} &= \mu - \mu M_{X}\left( \log(1-\alpha^{\ast})\right), \\
&= \mu - \mu \left( 1-\alpha^{\ast} \right)^D.
\end{align}
We let $\sigma = \frac{\alpha^{\ast}}{\mu}$. Then $\sigma$ is the solution to
\begin{equation}
\label{eq:k2}
\sigma = 1 - \left( 1 - \mu\sigma\right)^D.
\end{equation}
Using~\eqref{eq:k2}, and the fact that $M_{X}^{'}(t) = De^{Dt}$, in Theorem~\ref{thm:GGeo1} we obtain the result.

\begin{IEEEbiography}[{\includegraphics[width=1in,height=1.25in,clip,keepaspectratio]{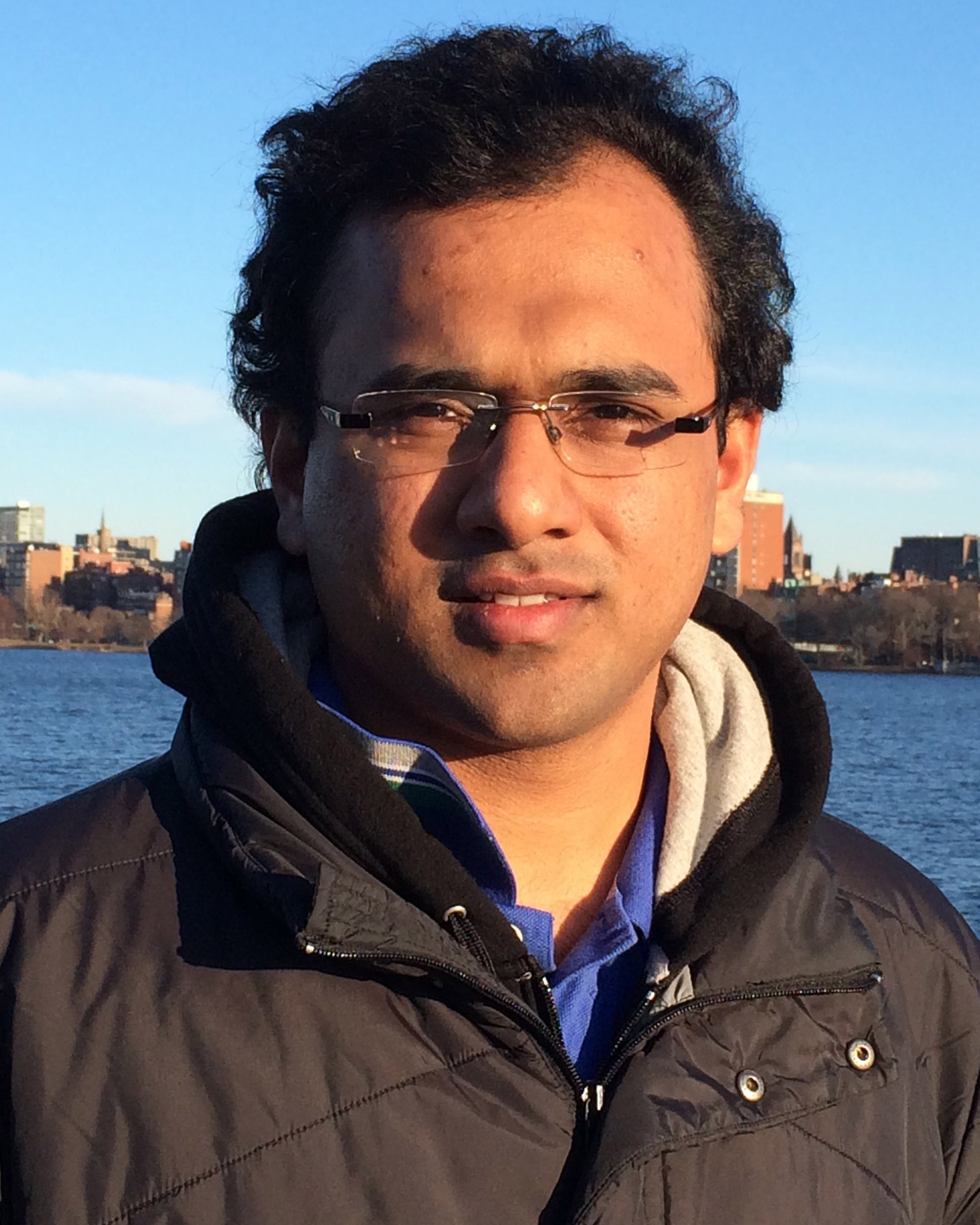}}]{Rajat Talak} is a PhD student in the Laboratory of Information and Decision Systems (LIDS) at Massachusetts Institute of Technology. He received a Master of Science degree from the Dept. of Electrical Communication Engineering, Indian Institute of Science, Bangalore, India, in 2013. He was awarded the prestigious Prof. F M Mowdawalla medal for his masters thesis. 

His research inclination is towards modeling, analysis, and design of algorithms for networked systems. His current focus has been towards studying age of information, and guaranteeing information freshness in wireless networks. His other notable work includes developing a new theoretical framework of uncertainty variables for machine learning and robotic perception. He is the recipient of ACM MobiHoc 2018 best paper award.
\end{IEEEbiography}

\begin{IEEEbiography}[{\includegraphics[width=1in,height=1.25in,clip,keepaspectratio]{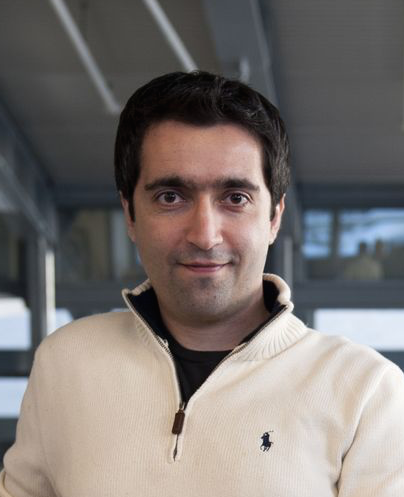}}]{Sertac Karaman} is an Associate Professor of Aeronautics and Astronautics at the Massachusetts Institute of Technology (since Fall 2012). He has obtained B.S. degrees in mechanical engineering and in computer engineering from the Istanbul Technical University, Turkey, in 2007; an S.M. degree in mechanical engineering from MIT in 2009; and a Ph.D. degree in electrical engineering and computer science also from MIT in 2012. His research interests lie in the broad areas of robotics and control theory. In particular, he studies the applications of probability theory, stochastic processes, stochastic geometry, formal methods, and optimization for the design and analysis of high-performance cyber-physical systems. The application areas of his research include driverless cars, unmanned aerial vehicles, distributed aerial surveillance systems, air traffic control, certification and verification of control systems software, and many others. He delivered the the Robotics: Science and Systems Early Career Spotlight Talk in 2017. He is the recipient of an IEEE Robotics and Automation Society Early Career Award in 2017, an Office of Naval Research Young Investigator Award in 2017, Army Research Office Young Investigator Award in 2015, National Science Foundation Faculty Career Development (CAREER) Award in 2014, AIAA Wright Brothers Graduate Award in 2012, and an NVIDIA Fellowship in 2011. He serves as the technical area chair for the Transactions on Aerospace Electronic Systems for the robotics area, a co-chair of the IEEE Robotics and Automation Society Technical Committee of Algorithms for the Planning and Control of Robot Motion.
\end{IEEEbiography}

\begin{IEEEbiography}[{\includegraphics[width=1in,height=1.25in,clip,keepaspectratio]{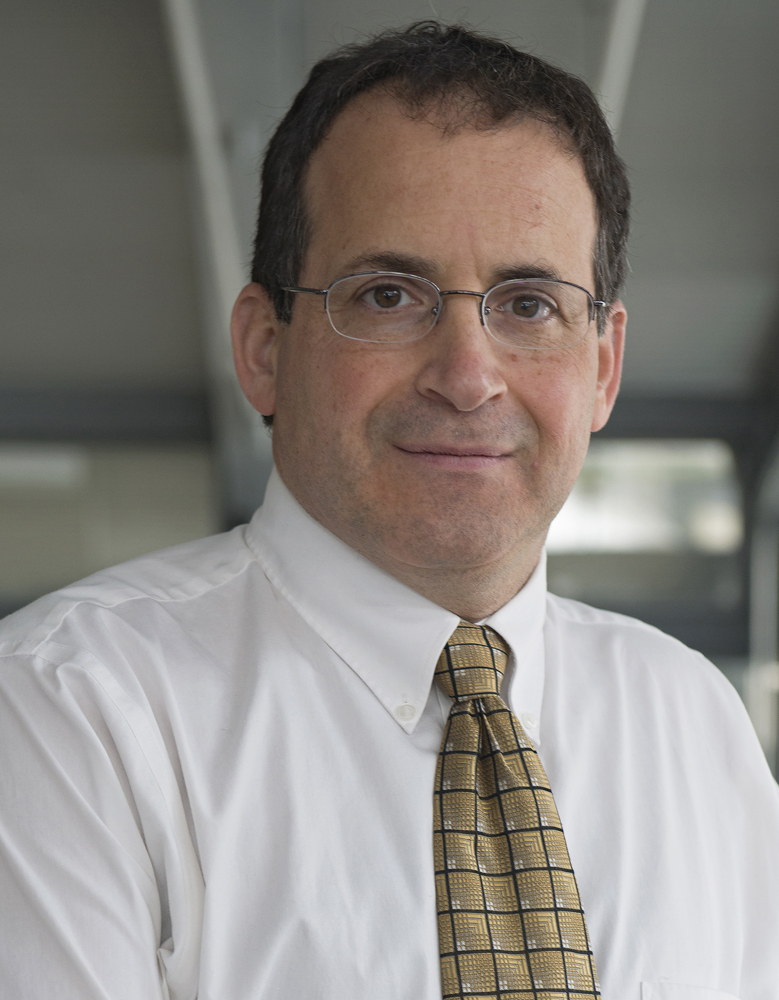}}]{Eytan Modiano} is Professor in the Department of Aeronautics and Astronautics and Associate Director of the Laboratory for Information and Decision Systems (LIDS) at MIT.  Prior to Joining the faculty at MIT in 1999, he was a Naval Research Laboratory Fellow between 1987 and 1992, a National Research Council Post Doctoral Fellow during 1992-1993, and a member of the technical staff at  MIT Lincoln Laboratory between 1993 and 1999.  Eytan Modiano received his B.S. degree in Electrical Engineering and Computer Science from the University of Connecticut at Storrs in 1986 and his M.S. and PhD degrees, both in Electrical Engineering, from the University of Maryland, College Park, MD, in 1989 and 1992 respectively.

His research is on modeling, analysis and design of communication networks and protocols.   He is the co-recipient of the Infocom 2018 Best paper award, the MobiHoc 2018 best paper award, the MobiHoc 2016 best paper award, the Wiopt 2013 best paper award, and the Sigmetrics 2006 best paper award.  He is the Editor-in-Chief for IEEE/ACM Transactions on Networking, and served as Associate Editor for IEEE Transactions on Information Theory and IEEE/ACM Transactions on Networking.  He was the Technical Program co-chair for  IEEE Wiopt 2006, IEEE Infocom 2007, ACM MobiHoc 2007, and DRCN 2015.  He had served on the IEEE Fellows committee in 2014 and 2015, and is a Fellow of the IEEE and an Associate Fellow of the AIAA.
\end{IEEEbiography}

\pagebreak


\section*{Supplementary material}

# Optimizing Information Freshness in Wireless Networks under General Interference Constraints

Rajat Talak, Sertac Karaman, Eytan Modiano

APPENDIX

*F. Proof of Lemma 2*

Let $X_n$ be the inter-arrival time between the $(n-1)$th and $n$th update packet at the queue, and $N_n$ be the number of update packets in the queue at the arrival of the $n$th update packet. If $Z_{n+1}$ be the number of service times we can fit in the duration $X_{n+1}$, then we have

$$N_{n+1} = \max\{N_n + 1 - Z_{n+1}, 0\}. \tag{1}$$

For this recursion, it is known that the steady state distribution of $N_n$ is geometric [1], i.e.,

$$\mathbf{P}[N = k] = \sigma(1-\sigma)^k, \tag{2}$$

for all $k \in \{0, 1, 2, \ldots\}$.

Now, the system time for the $n$th update packet is given by

$$T_n = \sum_{j=1}^{N_n+1} S_j, \tag{3}$$

where $S_j$ are independent, geometrically distributed random variables over $\{1, 2, \ldots\}$ with rate $\mu$. Since $N_n$ is geometrically distributed over $\{0, 1, 2, \ldots\}$ at steady state, $N_n + 1$ will also be geometrically distributed over $\{1, 2 \ldots\}$ at steady state, with the same rate $\sigma$. With $N_n + 1$ and $S_j$ being independent and geometrically distributed, we have from (3) that $T_n$ is also geometrically distributed over $\{1, 2, \ldots\}$ with rate $\sigma\mu$.

Let $\alpha = \sigma\mu$ be the rate of geometrically distributed random variable $T_n$ at steady state. We obtain (21) using the system time recursive equation. We know that the system time $T_n$ follows the recursive equation:

$$T_n = \max\{T_{n-1} - X_n, 0\} + S_n, \tag{4}$$

where $S_n$ is the service time of the $n$th update packet. Taking expected values on both sides we get

$$\frac{1}{\alpha} = \mathbb{E}\left[\max\{T_{n-1} - X_n, 0\}\right] + \frac{1}{\mu}, \tag{5}$$

because $T_n$ and $S_n$ are geometrically distributed random variables over $\{1, 2, \ldots\}$ with rates $\alpha$ and $\mu$, respectively. The quantity $\mathbb{E}\left[\max\{T_{n-1} - X_n, 0\}\right]$ can be computed as follows:

$$\begin{aligned}\mathbb{E}&\left[\max\{T_{n-1} - X_n, 0\}\right] \\ &= \mathbb{E}\left[\mathbb{E}\left[\max\{T_{n-1} - X_n, 0\}|X_n\right]\right], \\ &= \sum_{m=1}^{\infty} \mathbf{P}\left[X_n = m\right] \mathbb{E}\left[\max\{T_{n-1} - m, 0\}\right],\end{aligned} \tag{6}$$

where the last equality follows from the fact that the system time for the $(n-1)$th update packet $T_{n-1}$ and the inter-generation time between $(n-1)$th and $n$th update packet $X_n$ are independent. Using the fact that, at steady state, $T_{n-1}$ is geometrically distributed over $\{1, 2, \ldots\}$ with rate $\alpha$, we get

$$\begin{aligned}\mathbb{E}\left[\max\{T_{n-1} - m, 0\}\right] &= \sum_{k=1}^{\infty} k\alpha(1-\alpha)^{m+k-1}, \\ &= (1-\alpha)^m \sum_{k=1}^{\infty} k\alpha(1-\alpha)^{k-1}, \\ &= \frac{(1-\alpha)^m}{\alpha}. \end{aligned} \tag{7}$$

Substituting (7) in (6) we get

$$\begin{aligned}\mathbb{E}&\left[\max\{T_{n-1} - X_n, 0\}\right] \\ &= \frac{1}{\alpha} \sum_{m=1}^{\infty} \mathbf{P}\left[X_n = m\right](1-\alpha)^m, \\ &= \frac{1}{\alpha} M_X(\log(1-\alpha)). \end{aligned} \tag{8}$$

Substituting (8) back in (5) we obtain the result of (21).

*G. Proof of Theorem 5*

The average age is given by [2]

$$A^{\text{ave}} = \lambda\left[\frac{1}{2}\mathbb{E}\left[X_n^2\right] + \mathbb{E}\left[T_n X_n\right]\right] + \frac{1}{2}, \tag{9}$$

where $T_n$ is the system time for the $n$th update packet at steady state, and $X_n$ is the inter-generation time between $(n-1)$th and $n$th update packets. It is clear that $\mathbb{E}\left[X_n^2\right] = M_X''(0)$. Therefore, it suffices to show that

$$\lambda\mathbb{E}\left[T_n X_n\right] = \frac{\lambda}{\alpha^*} M_X'(\log(1-\alpha^*)) + \frac{1}{\mu}, \tag{10}$$

at steady state, for $\alpha^*$ given in Lemma 2.

We know that the system time follows the following recursive equation [1]:

$$T_n = \max\{T_{n-1} - X_n, 0\} + S_n, \tag{11}$$

where $S_n$ is the service time of the $n$th update packet, and is geometrically distributed over $\{1, 2, \ldots\}$ with rate $\mu$. Therefore, $\mathbb{E}[T_n X_n]$ can be computed as

$$\mathbb{E}[T_n X_n] = \mathbb{E}[\mathbb{E}[T_n X_n | X_n]],$$
$$= \sum_{m=1}^{\infty} m \mathbb{E}[T_n | X_n = m] \mathbf{P}[X_n = m]. \quad (12)$$

Now, $\mathbb{E}[T_n | X_n = m]$ can be evaluated using the recursion in (11) as

$$\mathbb{E}[T_n | X_n = m] = \mathbb{E}[\max\{T_{n-1} - m, 0\} + S_n | X_n = m],$$
$$= \mathbb{E}[\max\{T_{n-1} - m, 0\}] + \mathbb{E}[S_n],$$

where the last equality follows because the service time $S_n$ is independent of inter-generation time of update packets $X_n$. Since $\mathbb{E}[S_n] = \frac{1}{\mu}$ we get

$$\mathbb{E}[T_n | X_n = m] = \mathbb{E}[\max\{T_{n-1} - m, 0\}] + \frac{1}{\mu},$$
$$= \sum_{k=1}^{\infty} k \alpha^* (1 - \alpha^*)^{k+m-1} + \frac{1}{\mu}, \quad (13)$$
$$= (1 - \alpha^*)^m \sum_{k=1}^{\infty} k \alpha^* (1 - \alpha^*)^{k-1} + \frac{1}{\mu},$$
$$= \frac{1}{\alpha^*} (1 - \alpha^*)^m + \frac{1}{\mu}, \quad (14)$$

where (13) follows because at steady state, $T_{n-1}$ is geometrically distributed over $\{1, 2, \ldots\}$ with rate $\alpha^*$; see (7) in Appendix F. Substituting (14) in (12) we obtain

$$\mathbb{E}[T_n X_n] = \sum_{m=1}^{\infty} m \left( \frac{1}{\alpha^*} (1 - \alpha^*)^m + \frac{1}{\mu} \right) \mathbf{P}[X_n = m],$$
$$= \frac{1}{\alpha^*} M'_X(\log(1 - \alpha^*)) + \frac{1}{\lambda} \frac{1}{\mu}, \quad (15)$$

where the last equality follows from

$$M'_X(\log(1 - \alpha^*)) = \sum_{m=1}^{\infty} m (1 - \alpha^*)^m \mathbf{P}[X_n = m], \quad (16)$$

which can be obtained using standard properties of moment generating function. This proves (10), and therefore, the result follows.

### H. Derivation of Peak and Average Age for Ber/Ber/1 Queue

When update packet generation is Bernoulli with rate $\lambda$, the inter-generation time $X$ is geometrically distributed with rate $\lambda$. Thus, the moment generating function $M_X$ is given by [3]

$$M_X(t) = \frac{\lambda e^t}{1 - (1 - \lambda) e^t}. \quad (17)$$

Therefore, (21) of Lemma 2 becomes

$$\alpha = \mu - \mu M_X(\log(1 - \alpha)),$$
$$= \mu - \mu \left( \frac{\lambda(1 - \alpha)}{1 - (1 - \lambda)(1 - \alpha)} \right),$$

which upon solving for $\alpha$ yields

$$\alpha^* = \frac{\mu - \lambda}{1 - \lambda}. \quad (18)$$

**Peak Age**: Substituting this $\alpha^*$, given by (18), in (22) of Theorem 5 we get peak age to be

$$A^{\text{p}} = \frac{1}{\alpha^*} + \frac{1}{\lambda} = \frac{1 - \lambda}{\mu - \lambda} + \frac{1}{\lambda},$$

which can be written as

$$A^{\text{p}} = \frac{1}{\mu} \left[ \frac{1}{1 - \rho} + \frac{1}{\rho} \right] - \frac{\rho}{1 - \rho}, \quad (19)$$

where $\rho = \frac{\lambda}{\mu}$.

**Average Age**: To compute average age using Theorem 5, need to obtain expressions for $M''_X(0)$, which is $\mathbb{E}[X^2]$, and $M'_X(\log(1 - \alpha^*))$. Given that $X$ is geometrically distributed random variable over $\{1, 2, \ldots\}$ with rate $\lambda$, we know that [3]

$$M''_X(0) = \mathbb{E}[X^2] = \frac{2 - \lambda}{\lambda^2}. \quad (20)$$

Furthermore,

$$M'_X(t) = \frac{\lambda e^t}{(1 - (1 - \lambda) e^t)^2}, \quad (21)$$

and thus,

$$M'_X(\log(1 - \alpha^*)) = \frac{\lambda(1 - \alpha^*)}{(1 - (1 - \lambda)(1 - \alpha^*))^2}. \quad (22)$$

Substituting $1 - \alpha^* = \frac{1 - \mu}{1 - \lambda}$ from (18) in (22) we obtain

$$M'_X(\log(1 - \alpha^*)) = \frac{\lambda}{\mu^2} \frac{1 - \mu}{1 - \lambda}. \quad (23)$$

Using (18), (20) and (23) in (23) of Theorem 5 we obtain the result.

### I. Proof of Lemma 3

The service rate at link $e$ is $\mu_e = \gamma_e f_e$. The peak and average age functions, namely $A_e^{\text{p}}$ in (25) and $A_e^{\text{ave}}$ in (26), depend on the link activation frequency $f_e$, only via the product $\mu_e = \gamma_e f_e$. It, therefore, suffices to prove the result (33) and (34), for all $\mu_e \in (0, 1)$. For the ease of presentation, we drop the subscript $e$ in the following, and use $\rho$ and $\mu = \gamma f$ instead of $\rho_e$ and $\mu_e = \gamma_e f_e$, respectively.

Define

$$B(\mu, \rho) = \frac{1}{\mu} [H(\rho) + G(\rho)] - G(\rho), \quad (24)$$

for $\rho \in (0, 1]$, $\mu \in (0, 1]$. The functions $H(\rho)$ and $G(\rho)$, defined over the domain $(0, 1]$, satisfy the following assumptions:

A1) $H(\rho)$ is strictly decreasing, strongly convex, and twice differentiable.
A2) $H(\rho) \to +\infty$ as $\rho \to 0$, whereas $0 < H(1) < +\infty$.
A3) $G(\rho)$ is strictly increasing, strongly convex, and twice differentiable.
A4) $G(\rho) \to +\infty$ as $\rho \to 1$, whereas $G(0) < +\infty$.

For the function $F(\rho) = H(\rho) + G(\rho)$, define
$$\hat{\rho} = \arg \min_{\rho \in (0,1]} F(\rho), \quad (25)$$
and
$$\rho(\mu) = \arg \min_{\rho \in (0,1]} B(\mu, \rho). \quad (26)$$

It must be noted that both, $F(\rho)$ and $B(\mu, \rho)$, have unique minimizers as they are strictly convex over a bounded domain. Define the function difference $\Delta(\mu)$ to be
$$\Delta(\mu) = B(\mu, \hat{\rho}) - B(\mu, \rho(\mu)). \quad (27)$$

We prove the following bound on $\Delta(\mu)$ in Appendix J.

**Lemma 1:** $\Delta(\mu) \leq H(\hat{\rho}) - H(1)$.

Our result, for both peak and average age, follow directly from Lemma 1.

**Peak Age case:** Choosing $H(\rho) = 1 + \frac{1}{\rho}$ and $G(\rho) = \frac{\rho}{1-\rho}$ yields $B(\mu, \rho) = A^p(f, \rho)$ and $\hat{\rho} = \bar{\rho}^p = 1/2$; note that $\mu = \gamma f$. Using this, $\Delta(\mu)$ can be written as:

$$\Delta(\mu) = B(\mu, \bar{\rho}^p) - \min_{\rho \in (0,1]} B(\mu, \rho),$$
$$= A^p(f, \bar{\rho}^p) - \min_{\rho \in (0,1]} A^p(f, \rho), \quad (28)$$
$$= A^p(f, \bar{\rho}^p) - \min_{\rho \in (0,1]} A^p(f, \rho), \quad (29)$$
$$\leq H(\bar{\rho}^p) - H(1) = \frac{1}{\bar{\rho}^p} - 1 = 1, \quad (30)$$

where the last inequality follows by invoking the bound on $\Delta(\mu)$ in Lemma 1.

**Average Age case:** Choosing $H(\rho) = 1 + \frac{1}{\rho}$ and $G(\rho) = \frac{\rho^2}{1-\rho}$ yields $B(\mu, \rho) = A^{\text{ave}}(f, \rho)$ and $\hat{\rho} = \bar{\rho}^{\text{ave}} \approx 0.53$; note that $\mu = \gamma f$. Using this, $\Delta(\mu)$ can be written as:

$$\Delta(\mu) = B(\mu, \bar{\rho}^{\text{ave}}) - \min_{\rho \in (0,1]} B(\mu, \rho),$$
$$= A^{\text{ave}}(f, \bar{\rho}^{\text{ave}}) - \min_{\rho \in (0,1]} A^{\text{ave}}(f, \rho), \quad (31)$$
$$= A^{\text{ave}}(f, \bar{\rho}^{\text{ave}}) - \min_{\rho \in (0,1]} A^{\text{ave}}(f, \rho), \quad (32)$$
$$\leq H(\bar{\rho}^{\text{ave}}) - H(1) = \frac{1}{\bar{\rho}^{\text{ave}}} - 1 < 2 - 1 = 1, \quad (33)$$

since $\bar{\rho}^{\text{ave}} > 1/2$, where we have used Lemma 1 in the first inequality.

*J. Proof of Lemma 1*

Consider $B(\mu, \rho)$, $H(\rho)$, $G(\rho)$, $F(\rho)$, $\hat{\rho}$, $\rho(\mu)$, and $\Delta(\mu)$ as defined in Appendix I. We prove Lemma 1.

We first prove the following properties of $\rho(\mu)$:

**Lemma 2:**
1) $\rho(\mu)$ is non-decreasing and differentiable in $\mu$.
2) $\lim_{\mu \to 0} \rho(\mu) = \hat{\rho}$ and $\rho(1) = 1$.
3) $\hat{\rho} \leq \rho(\mu) \leq 1$.

*Proof:* Note that part 3 follows from the first two. It therefore suffices to prove only the first two statements.

**1.** Firstly, notice that $\rho(\mu) = \arg \min_{\rho \in (0,1]} H(\rho) + (1-\mu)G(\rho)$. The function $H(\rho) + (1-\mu)G(\rho)$ varies only linearly in $\mu$. As a result, the differentiability of $\rho(\mu)$ with respect to $\mu$ follows from [4]; we make use of the assumptions A1-A4 here. We know that $\rho(\mu)$ satisfies the first order condition $\frac{d}{d\rho}B(\mu,\rho)\big|_{\rho=\rho(\mu)} = 0$. This yields:

$$H'(\rho(\mu)) + G'(\rho(\mu)) = \mu G'(\rho(\mu)). \quad (34)$$

Differentiating this with respect to $\mu$ we obtain

$$\rho'(\mu) = \frac{G'(\rho(\mu))}{H''(\rho(\mu)) + (1-\mu)G''(\rho(\mu))} \geq 0, \quad (35)$$

where the inequality follows because $G$ is strictly increasing, $\mu \in (0,1]$, and both $G$ and $H$ are strongly convex. We, thus, conclude that $\rho(\mu)$ is an increasing function.

**2.** $\rho(1) = 1$ follows from the fact that $H(\rho)$ is a strictly decreasing function. We know that $\rho(\mu)$ is also continuous on $\mu \in [0,1]$. Note that $\rho(\mu) = \arg \min_\rho H(\rho) + (1-\mu)G(\rho)$ and $\bar{\rho} = \arg \min_\rho H(\rho) + G(\rho)$. Therefore, $\lim_{\mu \to 0} \rho(\mu) = \bar{\rho}$. ∎

If the difference function $\Delta(\mu)$ were to be non-decreasing then we can obtain a trivial upper-bound as follows:

$$\Delta(\mu) \leq \lim_{\mu \to 1} \Delta(\mu) = \lim_{\mu \to 1} B(\mu, \hat{\rho}) - \lim_{\mu \to 1} B(\mu, \rho(\mu)). \quad (36)$$

Noting that $\lim_{\mu \to 1} B(\mu, \hat{\rho}) = H(\hat{\rho})$ and $B(\mu, \rho(\mu)) \geq \frac{1}{\mu} H(\rho(\mu))$ we get

$$\Delta(\mu) \leq H(\hat{\rho}) - \lim_{\mu \to 1} \frac{1}{\mu} H(\rho(\mu)) = H(\hat{\rho}) - H(1). \quad (37)$$

It suffices to prove that $\Delta(\mu)$ is non-decreasing. We do so by showing that its first derivative to be non-negative:

$$\Delta'(\mu) = -\frac{1}{\mu^2}[F(\bar{\rho}) - F(\rho(\mu))] - \frac{1}{\mu}\left[F'(\rho(\mu))\rho'(\mu)\right]$$
$$+ G'(\rho(\mu))\rho'(\mu)$$
$$\geq 0, \quad (38)$$

because $F(\bar{\rho}) \leq F(\rho(\mu))$ as $\bar{\rho}$ minimizes $F(\rho)$, $F'(\rho) \leq 0$ since $F$ is decreasing, $\rho'(\mu) \geq 0$ and $G'(\rho) \geq 0$, as both are increasing function.

*K. Proof of Theorem 6*

*Peak age:* Using Lemma 3, we obtain

$$\sum_{e \in E} w_e A_e^p(f_e, \bar{\rho}^p) \leq \min_{\boldsymbol{\rho} \in [0,1]^{|E|}} \sum_{e \in E} w_e A_e^p(f_e, \rho_e) + \sum_{e \in E} w_e. \quad (39)$$

Minimizing over $\mathbf{f} \in \mathcal{F}$ we get

$$\underset{\mathbf{f} \in \mathcal{F}}{\text{Minimize}} \sum_{e \in E} w_e A_e^p(f_e, \bar{\rho}^p) \leq A_{\mathcal{B}}^{p*} + \sum_{e \in E} w_e, \quad (40)$$

since

$$A_{\mathcal{B}}^{p*} = \underset{\mathbf{f} \in \mathcal{F}, \boldsymbol{\rho} \in [0,1]^{|E|}}{\text{Minimize}} \sum_{e \in E} w_e A_e^p(f_e, \rho_e).$$

Now, notice that

$$\sum_{e \in E} w_e A_e^{\text{p}}(f_e, \bar{\rho}^{\text{p}}) = \left(\frac{1}{\bar{\rho}^{\text{p}}} + \frac{1}{1-\bar{\rho}^{\text{p}}}\right) \sum_{e \in E} \frac{w_e}{\gamma_e f_e}$$
$$- \left(\frac{\bar{\rho}^{\text{p}}}{1-\bar{\rho}^{\text{p}}}\right) \sum_{e \in E} w_e, \quad (41)$$

which has the same form as the objective function in (18), except for a constant multiple and a additive factor. As a result, stationary policy $\pi_C$ minimizes $\sum_{e \in E} w_e A_e^{\text{p}}(f_e, \bar{\rho}^{\text{p}})$. Thus, given $\mathbf{f}^*$ as the vector of link activation frequencies for policy $\pi_C$, we have

$$A^{\text{p}}(\mathbf{f}^*, \bar{\rho}^{\text{p}}\mathbf{1}) = \underset{\mathbf{f} \in \mathcal{F}}{\text{Minimize}} \sum_{e \in E} w_e A_e^{\text{p}}(f_e, \bar{\rho}^{\text{p}}). \quad (42)$$

Substituting (42) in (40), and using $\sum_{e \in E} w_e = 1$, gives the result. The proof for the average age follows the same line of argument.

*L. Proof of Corollary 3*

**Peak Age**: First, note that the minimum peak age for the buffered sources is lower bounded by the minimum peak age in active source case:

$$A^{\text{p}*} \leq A_{\mathcal{QB}}^{\text{p}*}. \quad (43)$$

This is because fresh information is generated at the beginning of every slot in the active source case. Using (25), we can obtain the peak age $A^{\text{p}}(\mathbf{f}^*, \bar{\rho}^{\text{p}}\mathbf{1})$ of the SPP to be

$$A^{\text{p}}(\mathbf{f}^*, \bar{\rho}^{\text{p}}\mathbf{1}) = \left(\frac{1}{\bar{\rho}^{\text{p}}} + \frac{1}{1-\bar{\rho}^{\text{p}}}\right) \sum_{e \in E} \frac{w_e}{\gamma_e f_e^*} - \left(\frac{\bar{\rho}^{\text{p}}}{1-\bar{\rho}^{\text{p}}}\right), \quad (44)$$

where we have used $\sum_{e \in E} w_e = 1$. Noting that $A^{\text{p}*} = \sum_{e \in E} \frac{w_e}{\gamma_e f_e^*}$, we obtain

$$A^{\text{p}}(\mathbf{f}^*, \bar{\rho}^{\text{p}}\mathbf{1}) \leq \left(\frac{1}{\bar{\rho}^{\text{p}}} + \frac{1}{1-\bar{\rho}^{\text{p}}}\right) A^{\text{p}*}$$
$$\leq \left(\frac{1}{\bar{\rho}^{\text{p}}} + \frac{1}{1-\bar{\rho}^{\text{p}}}\right) A_{\mathcal{QB}}^{\text{p}*}, \quad (45)$$

which follows from (43). This yields the result.

**Average Age**: Just as for the peak age, the minimum average age for the buffered sources is lower bounded by the minimum average age for the active source, i.e.,

$$A^{\text{ave}*} \leq A_{\mathcal{QB}}^{\text{ave}*}. \quad (46)$$

This is because fresh packets are available at every slot in the active source case. Using Corollary 1 and (46) we get

$$A^{\text{p}*} \leq 2 A_{\mathcal{QB}}^{\text{ave}*} - 1. \quad (47)$$

Using the upper-bound in (30) we get

$$A^{\text{ave}}(\mathbf{f}^*, \bar{\rho}^{\text{ave}}\mathbf{1}) \leq \left(1 + \frac{1}{\bar{\rho}^{\text{ave}}} + \frac{(\bar{\rho}^{\text{ave}})^2}{1-\bar{\rho}^{\text{ave}}}\right) \sum_{e \in E} \frac{w_e}{\gamma_e f_e^*}, \quad (48)$$

$$= \left(1 + \frac{1}{\bar{\rho}^{\text{ave}}} + \frac{(\bar{\rho}^{\text{ave}})^2}{1-\bar{\rho}^{\text{ave}}}\right) A^{\text{p}*}, \quad (49)$$

$$\leq 2\left(1 + \frac{1}{\bar{\rho}^{\text{ave}}} + \frac{(\bar{\rho}^{\text{ave}})^2}{1-\bar{\rho}^{\text{ave}}}\right) A_{\mathcal{QB}}^{\text{ave}*}, \quad (50)$$

where the last inequality follows from (47).



\end{document}